\documentclass[twocolumn,10pt,aps,longbibliography]{revtex4-2}
\usepackage{amsmath,amssymb,amsfonts}
\usepackage{algorithmic}
\usepackage{graphicx}
\usepackage{textcomp}
\usepackage{float}
\usepackage[caption = false]{subfig}
\usepackage{enumitem}
\usepackage{multirow}
\usepackage{array}
\newcolumntype{M}[1]{>{\centering\arraybackslash}m{#1}}
\RequirePackage{filecontents}
\setlist{nolistsep}
\usepackage[separate-uncertainty = true]{siunitx}
\def\BibTeX{{\rm B\kern-.05em{\sc i\kern-.025em b}\kern-.08em
    T\kern-.1667em\lower.7ex\hbox{E}\kern-.125emX}}
\PassOptionsToPackage{hyphens}{url}\usepackage{hyperref}

\begin{document}

\title{Comparison of Techniques for Distillation of Entanglement over a Lossy Channel}
\author{Caroline Mauron and Timothy C. Ralph}
\affiliation{Centre for Quantum Computation and Communication Technology, School of Mathematics and Physics, University of Queensland, St Lucia, Queensland 4072, Australia}

\begin{abstract}

We analyze three quantum communication protocols that have been proposed in the literature, and compare how well they communicate single-rail entanglement. We use specific metrics for output state purity and probability of success and include the presence of imperfect photon source and detection components. We find that a distributed noiseless linear amplification (NLA) protocol with a relay point placed half-way between Alice and Bob outperforms NLA at Bob's end and a recently proposed purification protocol under most conditions, unless the distance is very small or the photon source component is very good.

\end{abstract}

\maketitle

\section{Introduction}
\label{sec:introduction}

The distribution of entangled states is a critical resource in many quantum communication protocols \cite{b2, b3,b11}. Due to relatively low interaction of quantum states of light with their environment, single-photon entanglement and linear optics produce one of the simplest instances of such protocols \cite{b14}.

Nevertheless, practical implementation of these schemes has proven highly challenging due to inevitable losses in optical fibre transmission channels. Methods that were successful in overcoming optical loss for classical communication are unfortunately inadequate in the context of quantum communication. Indeed, it can be shown that deterministic amplification rapidly destroys entanglement, an effect that is not relevant for classical communication \cite{b4}.

Whilst historically most schemes have employed dual-rail encoding \cite{b18}, interest has recently increased in single-rail encoding \cite{b19}, where vacuum represent logical zero and one photon states represent logical one. In this paper we will focus on the distribution of single-rail entanglement.

Among various techniques that have been proposed to mitigate such loss, noiseless linear amplification (NLA) is an entanglement distillation protocol where the measurement result heralds whether the amplification was successful or not \cite{b5}. It has been used to demonstrate various loss mitigation schemes \cite{b20,b21,b22}. In particular, this protocol was experimentally demonstrated to be capable of distributing high-fidelity entanglement over loss-equivalent distances of up to 50 km \cite{b6}, and was subsequently used to implement an error corrected quantum channel \cite{b23}. A similar scheme with photon detection implemented at a central station half-way between Alice and Bob was also demonstrated experimentally \cite{b15}. This is a specific example of a general type of repeater protocol based on distributed NLA \cite{b17}. While NLA protocols produce entangled states with arbitrarily high but finite purity, purification protocols have also been proposed \cite{b24, b13} as a scheme to obtain perfect entangled state purity, at the cost of introducing additional source and detection components as well as a decreased probability of protocol success.

A key difficulty in the practical implementation of all such schemes is the efficiency of the single-photon source and detectors used by the protocol. Implementing such components in practice is technologically highly challenging. In recent state-of-the-art experiments, single photon detection was achieved with $98\%$ efficiency and high time-resolution using superconducting nanowire single photon detectors (SNSPDs) \cite{b9}. Furthermore, single-photon generation was recently demonstrated to yield a 66.7(24)\% probability of collecting a single photon using heralded single-photon sources (HSPS) \cite{b10, b16}. The tradeoff between state purity and heralded probability is therefore analysed here in the presence of these constraints.

We begin by introducing the NLA and our various assumptions in the next section. We illustrate our figures of merit initially by examining direct transmission of single-rail entanglement through a lossy channel. We then see how the situation changes when we introduce our two different NLA protocols. The purification protocol is then introduced in Section \ref{sec:purification}, and then we proceed to compare their performance in Section \ref{sec:comparison} before briefly concluding in Section \ref{sec:conclusion}.

\section{NLA Protocol}
\label{sec:analysis}

We first assume that perfect single-photon sources and single-photon detection devices are available, and later relax this assumption in subsection \ref{subsec:nlanoise}. Environmental noise is modelled as a vacuum state entering the open port of a beam splitter with transmissivity $\eta$, such that a photon sent by Alice to Bob either gets transmitted with probability $\eta$ or lost with probability $1-\eta$. A click or ``success'' is a measurement result that heralds entanglement of the state. In absence of a click, the state is discarded and another trial is made.

The aim of all protocols analyzed here is that whenever there is a click, Alice and Bob share a state whose entangled component is maximal (unbiased), and which contains as little environmental loss as possible. Such a final state can always be successfully targeted even if there is detection and/or source loss. We do not consider thermal noise of dark counts here, which could lead to additional error states if included.
Specifically, we define 
\begin{itemize}[noitemsep]
\item $P_{\text{success}}$ the probability of obtaining a click.
\item $\vert \psi \rangle$ one of the states that generate a click (with all such states typically sharing the same probability) 
\item $\vert \psi_f \rangle$ the projection of $\vert \psi \rangle$ on $_e \langle 0 \vert$ (ie, conditional on no environmental losses) 
\item $\vert \psi_0 \rangle$ the projection of $\vert \psi \rangle$ on $_e \langle 1 \vert$, or more generically for the other setup variations, on any environmental loss states.
\item $P_f$ and $P_0$ the respective probabilities of $\vert \psi_f \rangle$ and $\vert \psi_0 \rangle$ 
\item $X = \frac{P_f}{P_0}$ represents how pure the reduced state $\rho = \text{Tr}_e \left[ \vert \psi \rangle \langle \psi \vert \right]$ is, with $X=\infty$ for a perfectly pure state. The relationship of $X$ to more standard measures of purity such as $\text{Tr}\{\rho^2\}$ is $\text{Tr}\{\rho^2\} = (1+X^2)/(1+X)^2$.
\end{itemize}

\subsection{``Do Nothing'' Protocol}

In order to compare the performance of different protocols, a reference is first established by considering the metrics obtained in the absence of a distillation protocol. 

\begin{figure}[ht]
\centerline{\includegraphics[width=\columnwidth]{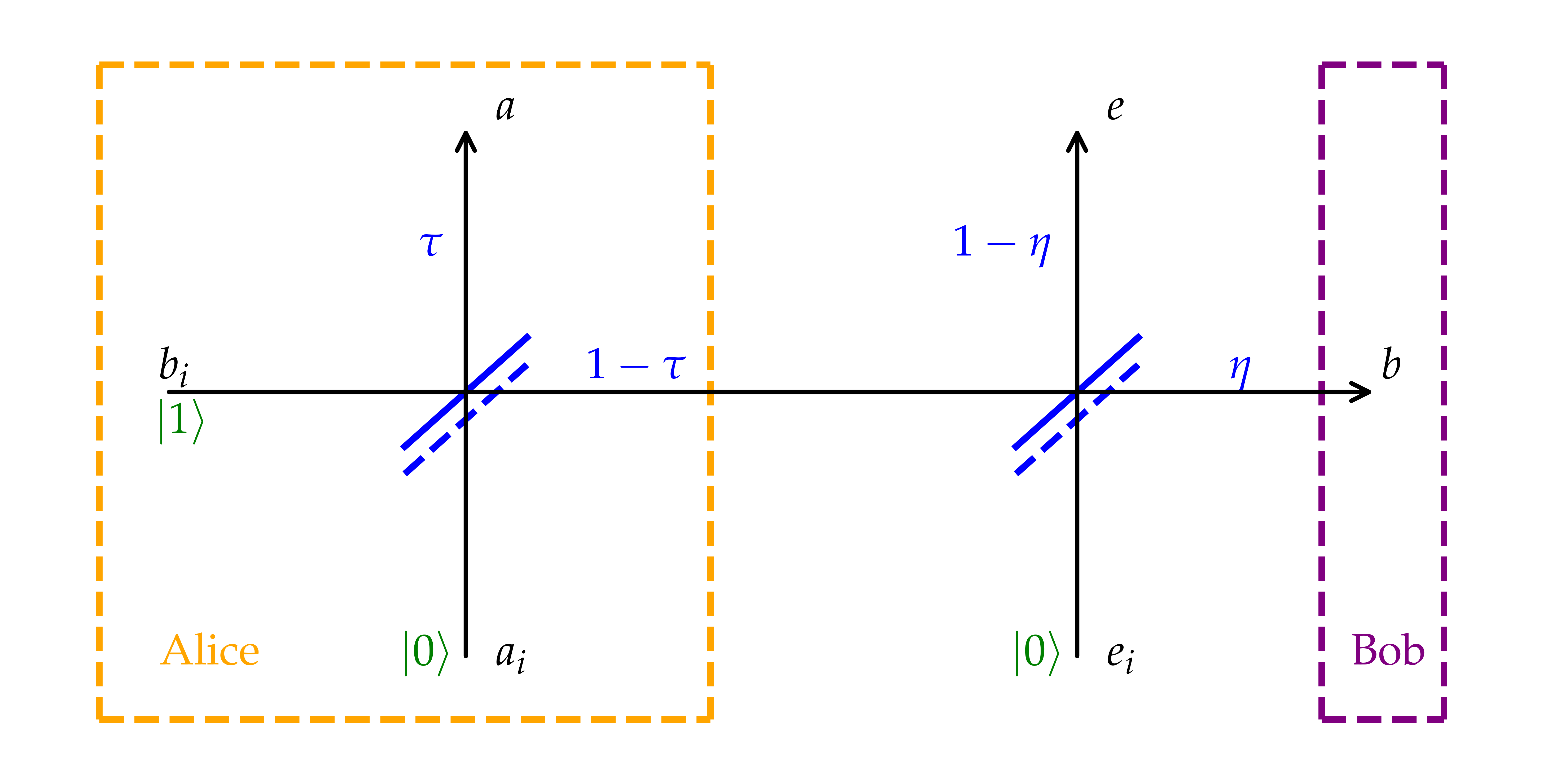}}
\caption{In the most basic framework, Alice prepares an entangled state using a beamsplitter then sends part of that state over to Bob through a lossy channel of transmissivity $\eta$. The channel loss is modelled as a vaccum state $\vert 0 \rangle_{e_i}$ entering through the open port of a beamsplitter. The parameter $\tau$ is chosen such that the resulting output state in modes $a$ and $b$ is a mixture of a pure maximally entangled state $\vert \psi_f \rangle$ and an environmental loss state $\vert \psi_0 \rangle$. The output state purity is quantified using the ratio $X = \frac{P_f}{P_0}$.}
\label{noprotocol}
\end{figure}

If Alice is in possession of a perfectly pure state with arbitrary entanglement $\tau$, and sends this state to Bob through the lossy channel of transmissivity $\eta$, as represented in Figure \ref{noprotocol}, the output state projected on no environmental losses is
     \begin{equation}
      \vert \psi_{f} \rangle  = \sqrt{\tau}  \vert 10 \rangle  _{ab}
  +   \sqrt{1-\tau}\sqrt{\eta} \vert 01 \rangle  _{ab} 
  \label{eq:dontg}
         \end{equation}
         with probability
     \begin{equation}
 P_f =  \tau  + \eta (1-\tau)
     \end{equation}
     while the projection on environmental losses produces
 \begin{equation}
 \vert \psi_{0} \rangle = \sqrt{1-\tau}\sqrt{1-\eta} \vert 00 \rangle  _{ab}  
\end{equation}
with probability
     \begin{equation}
 P_0 =   (1-\tau)(1-\eta)
 \end{equation}    
 From equation \ref{eq:dontg}, the condition of maximal entanglement yields 
      \begin{equation}
   \tau = \frac{\eta}{1 + \eta}
 \end{equation}   
 so that the state purity $X$ is
 \begin{equation}
 X=\frac{P_f}{P_0} = \frac{2 \eta}{1-\eta}
 \label{x_donothing}
 \end{equation}
 In this context, the protocol always succeeds, and $P_{\text{success}} = 1$.

\subsection{NLA at Bob's end}
\label{subsec:ref}

The building block of the NLA protocol as described and implemented in \cite{b5,b6} is illustrated in Figure \ref{diag1}. Details of the calculation are given in Appendix \ref{app:nla-perfect-ref}.

\begin{figure}[h]
\centerline{\includegraphics[width=\columnwidth]{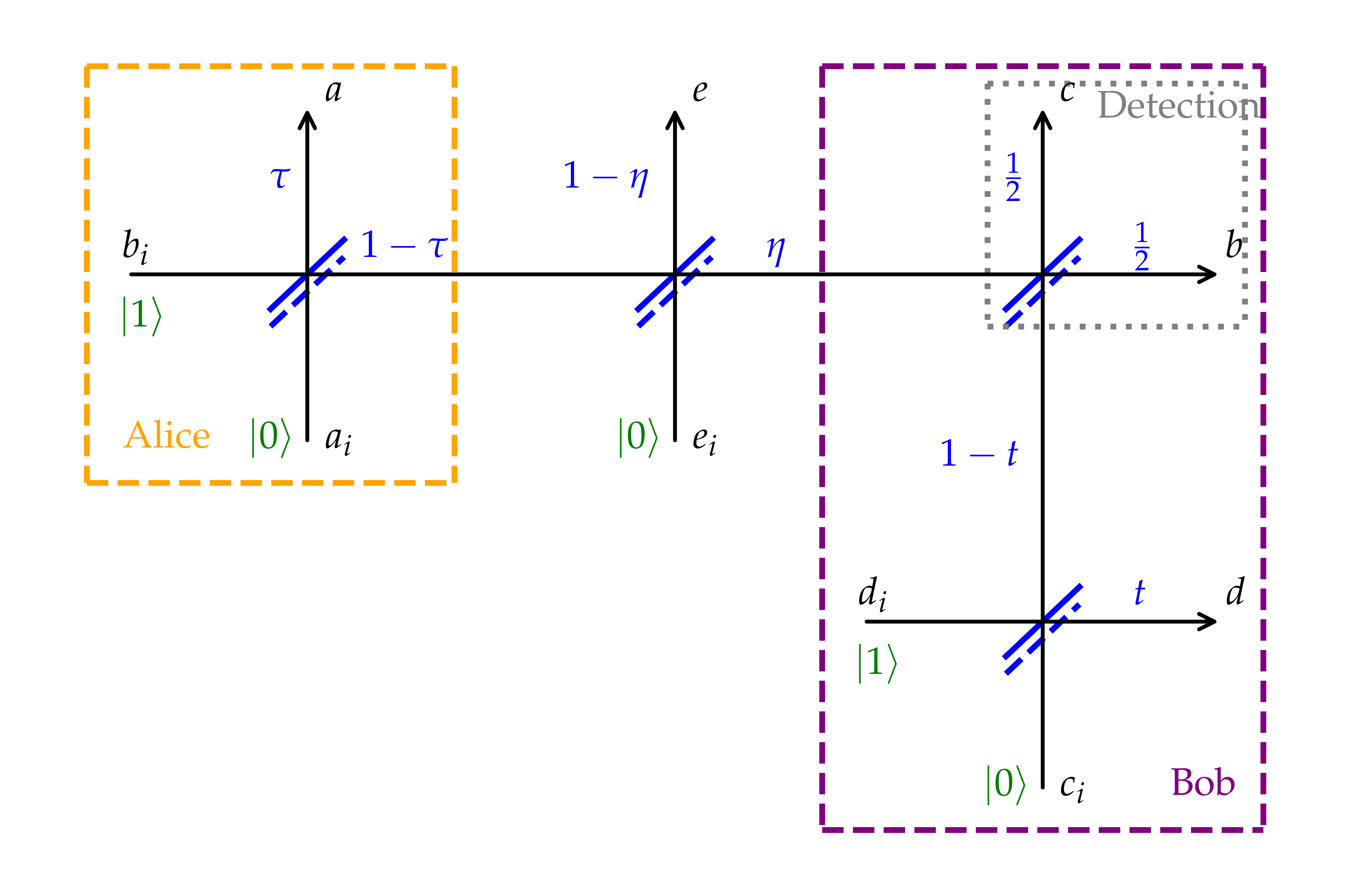}}
\caption{In the NLA protocol implemented at Bob's end, Alice and Bob each prepare an entangled state by sending a single photon through beamsplitters. Alice then sends part of her entangled state to Bob through a channel of transmissivity $\eta$. Bob interferes Alice's state and his own state through a 50:50 beamsplitter, and uses photon detection at the output modes $b$ and $c$. If exactly one photon is detected at mode $b$ or $c$ and no photon at the other mode (representing two possible click states), Alice and Bob will use their respective path-entangled states at the outputs $a$ and $d$.}
\label{diag1}
\end{figure}

We obtain 
     \begin{equation}
      \vert \psi_{f} \rangle  = \sqrt{\tau}  \sqrt{1-t}\sqrt{\frac{1}{2}} \vert 10 \rangle  _{ad}
  +   \sqrt{1-\tau}\sqrt{\eta} \sqrt{\frac{1}{2}}   \sqrt{t} \vert 01 \rangle  _{ad} 
         \end{equation}
         with probability
     \begin{equation}
 P_f = \langle \psi_{f} \vert \psi_{f} \rangle     =  \frac{1}{2}  \tau \left(1-t \right) +  \frac{1}{2}  \left( 1-\tau \right) \eta t
     \end{equation}
     and
 \begin{equation}
 \vert \psi_{0} \rangle =
\sqrt{1-\tau} \sqrt{1-\eta}   \sqrt{1-t} \sqrt{\frac{1}{2}} \vert 00 \rangle  _{ad}  
\label{psizero_orig_nonoise}
\end{equation}
with probability
     \begin{equation}
 P_0 =  \langle \psi_{0} \vert \psi_{0} \rangle     =   \frac{1}{2} \left(1-\tau \right)  \left(1-\eta \right)   \left(1-t \right) 
 \end{equation}    
 The condition of maximal entanglement yields 
      \begin{equation}
   \tau = \frac{t \eta}{1 - t + t \eta}
 \end{equation}   
 so that, after expressing $t$ in terms of $X$ and $\eta$, we get
\begin{equation}
 P_{\text{success}} = \frac{4 \eta  (1 - \eta) (1+X) } { \left( 2 \eta + X ( 1- \eta)  \right) \left( 2 + X ( 1- \eta)  \right)}
 \label{eq_psuc1}
\end{equation}

\begin{figure}[ht]
\centerline{\includegraphics[width=\columnwidth]{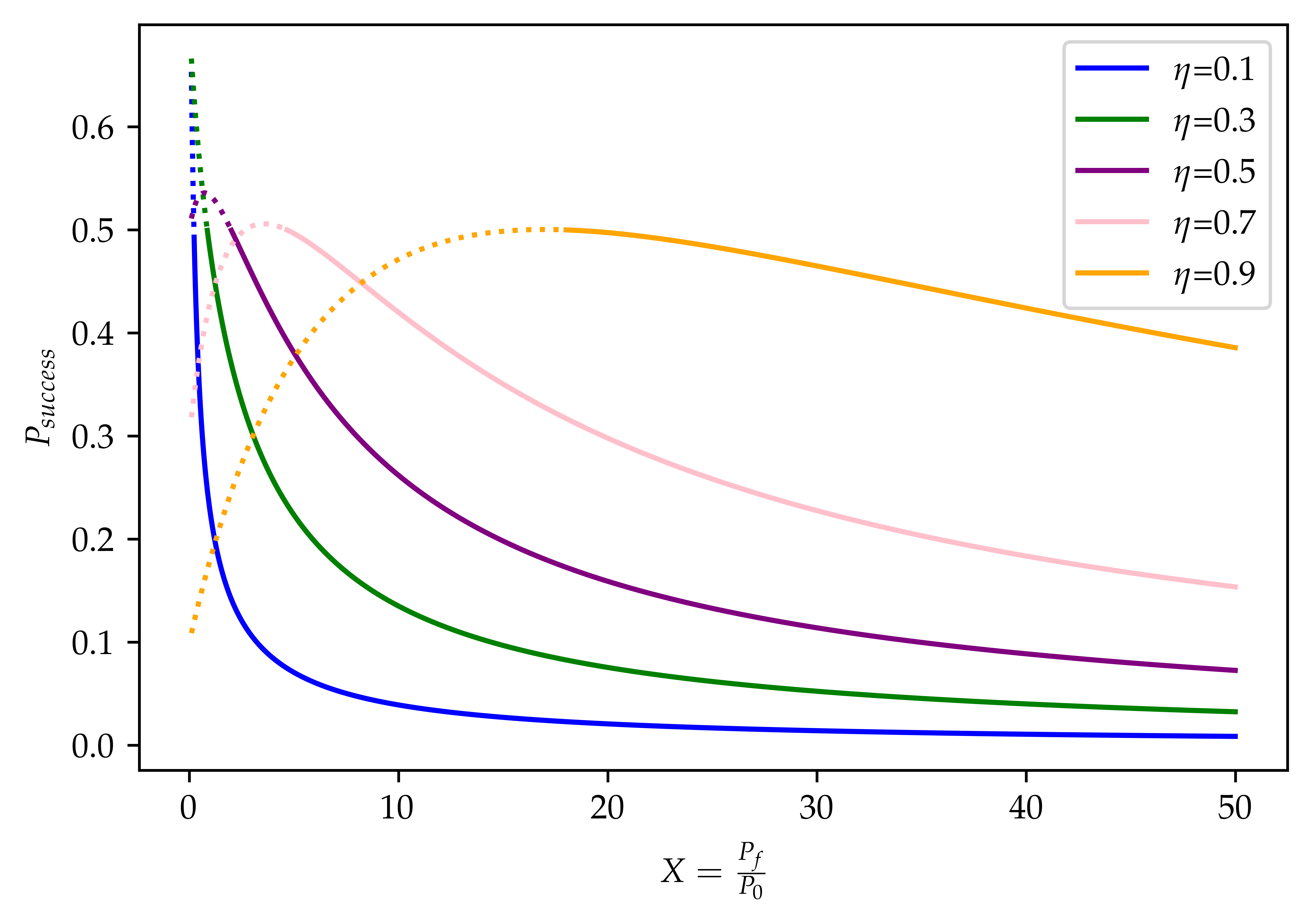}}
\caption{The probability $P_{\text{success}}$ of obtaining a click state is plotted as a function of the state purity $X$ for different channel transmissivities $\eta$. The dotted portion of the lines represent areas where a higher $X$ can be obtained from doing nothing, according to Equation \ref{x_donothing}. In the solid portion of the lines, the expected trade-off between click probability and state purity is observed, as the odds of obtaining a click decrease for higher purity.}
\label{psuc1}
\end{figure}

An important benefit of the NLA protocol lies in the fact that, given perfect source and detection components, the purity of the state $X$ can be made arbitrarily high by adjusting the entanglement of Bob's input state. This however is expected to come at the expense of a decreased click probability, a relationship illustrated in Figure \ref{psuc1}.

Surprisingly, we find that at small distances (high $\eta$, low losses), the click probability can actually increase with $X$. However,  as shown in Appendix \ref{app:nla-perfect-ref}, whenever this behaviour is observed, the $X$ produced by the NLA protocol is lower than the $X$ obtained without any protocol. Since the probability of success is also lower than 1, applying such protocol would be worse than doing nothing. In the more typical case (longer distances, $\eta \ll 1$), the expected behaviour is observed and the probabibility of success decreases as an increasingly purer state is targetted.

\subsection{NLA Half-way}
\label{subsec:alt}

An alternative setup as implemented in \cite{b15} is now analyzed, with photon detectors placed half-way between Alice and Bob, instead of at Bob's end. This setup is an example of the more general distributed NLA protocol in \cite{b17}. The transmissivity of a channel decays exponentially with distance, yielding the conventional assumption \cite{b6,b7,b8} of
\begin{equation}
\eta = e^{-\frac{d}{22}}
\end{equation}
where the attenuation loss distance of $\SI{22}{\km}$ is converted from typical optical fiber loss of $\SI{0.2}{\decibel \per \km}$. The transmissivity on each half-distance of the channel can therefore be modelled as $\sqrt{\eta}$, such that the total transmissivity over the whole distance is still $\eta$. 

\begin{figure}[ht]
\centerline{\includegraphics[width=\columnwidth]{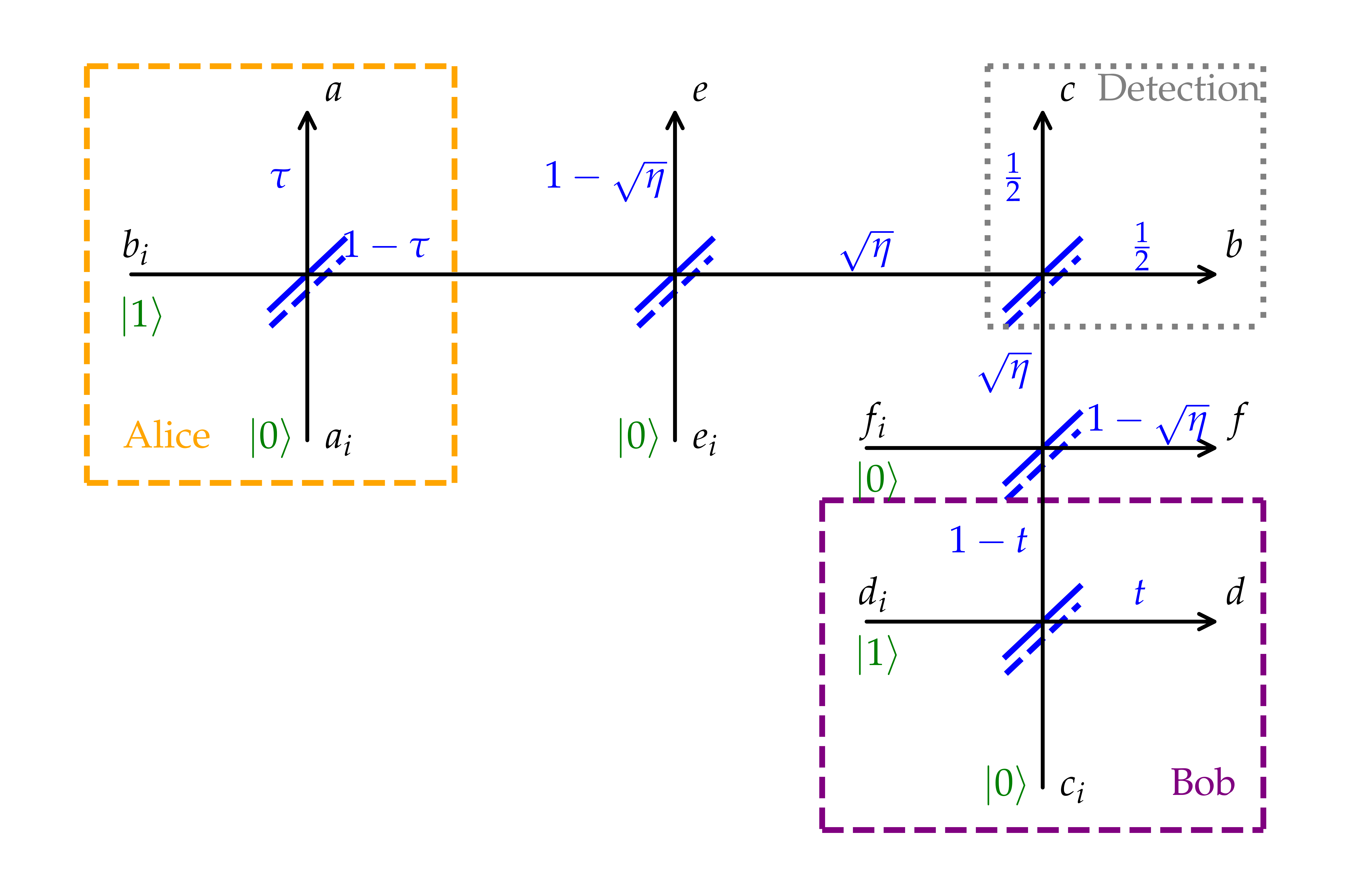}}
\caption{NLA diagram where the photon detection components $b$ and $c$ have been placed half-way between Alice and Bob. As a result, there are now two environmental loss modes $e$ and $f$, each modelled as a vacuum state entering the open port of a beamsplitter with transmissivity $\sqrt{\eta}$.}
\label{diag2}
\end{figure}

Repeating the calculations for the setup in Figure \ref{diag2}, the click probability is now
\begin{equation}
    P_{\text{success}}   = \frac{  2 \sqrt{\eta} (1-\sqrt{\eta}) (1+X)}{  (X (\sqrt{\eta} - 1) -1)^2  } 
 \label{eq_psuc2}
    \end{equation}
 (Key intermediary formulas are given in Appendix  \ref{app:nla-perfect-alt})
 
\begin{figure}[ht]
\centerline{\includegraphics[width=\columnwidth]{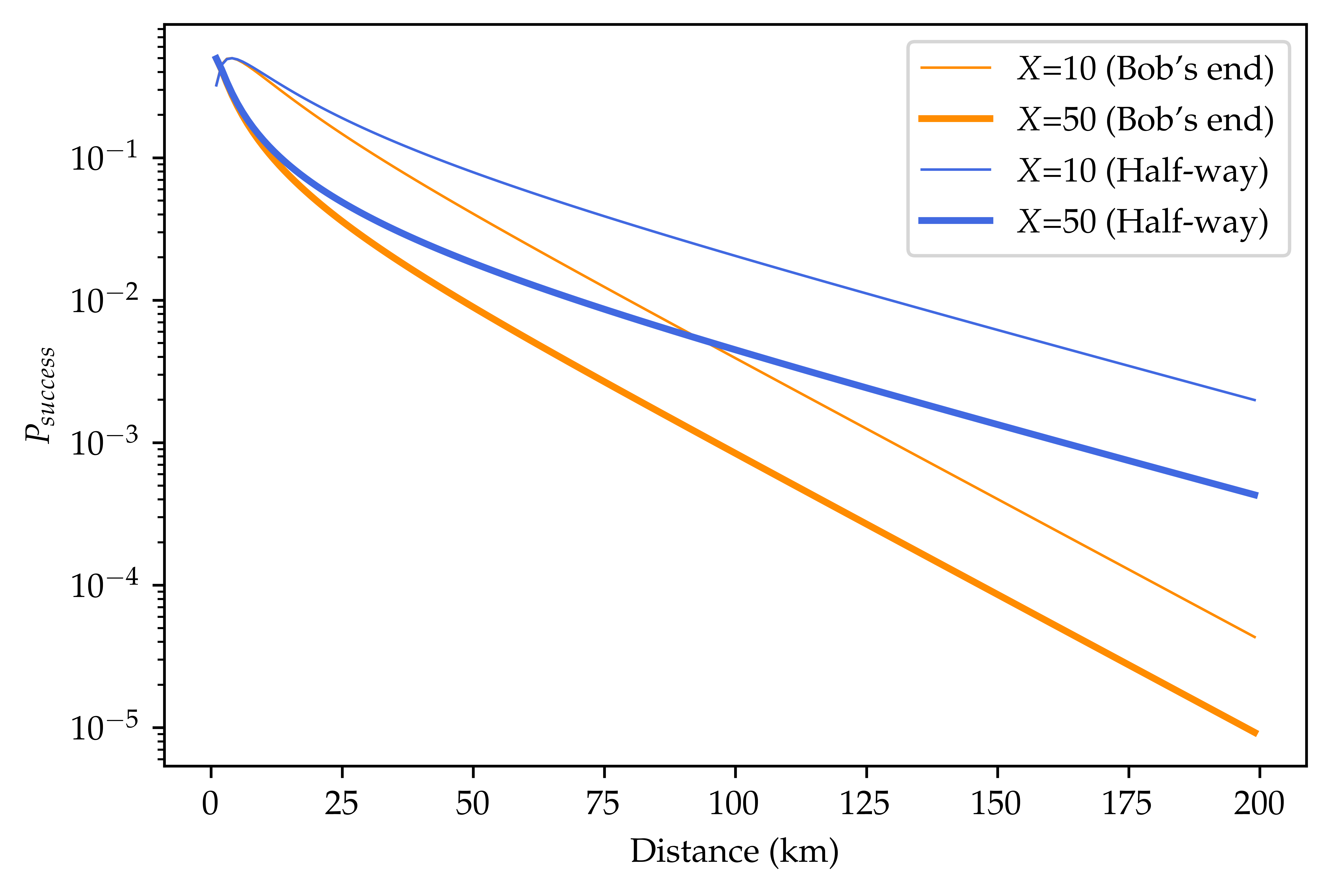}}
\caption{Comparison of the click probability in NLA protocols done either at Bob's end or half-way between Alice and Bob. At long-enough distances, the $\sqrt{\eta}$ scaling will always allow the half-way NLA protocol to outperform, regardless of the state purity chosen.}
\label{dist1}
\end{figure}

Comparing Equations \ref{eq_psuc1} and \ref{eq_psuc2}, we obtain the plot displayed in Figure \ref{dist1}. We see that for a given $ X \gg 1$ and assuming $\eta \ll 1$, the click probability in this alternative protocol will now scale with $\sqrt{\eta}$ instead of $\eta$, thus obtaining a much higher probability of success for a given state purity target.

\subsection{Addition of source and detection noise}
\label{subsec:nlanoise}

We now aim to compare the two setups (NLA at Bob's end and NLA half-way) in the presence of imperfect (i.e. inefficient) photon source and photon detection components. 
We analyse two situations: 
\begin{enumerate}[nolistsep,topsep=0pt,itemsep=-1ex,partopsep=1ex,parsep=1ex,label=(\roman*)]
\item perfect sources and imperfect detection (quality $\delta$) \label{sit1}
\item 1 imperfect source (quality $\epsilon$) on Bob's end and imperfect detection (quality $\delta$). \label{sit3}
\end{enumerate}

~

In each situation, we compare photon detection at Bob's end to photon detection half-way between Alice and Bob by going through the same calculations as in subsections \ref{subsec:ref} and \ref{subsec:alt}.
Imperfection at the source and the detectors are all modelled as additional environmental loss modes. The resulting diagrams for both situations are available in Appendices \ref{sec:app-nla-imp-detection} and \ref{sec:app-nla-imp-both}.

~

\begin{figure}[ht]
\centerline{\includegraphics[width=\columnwidth]{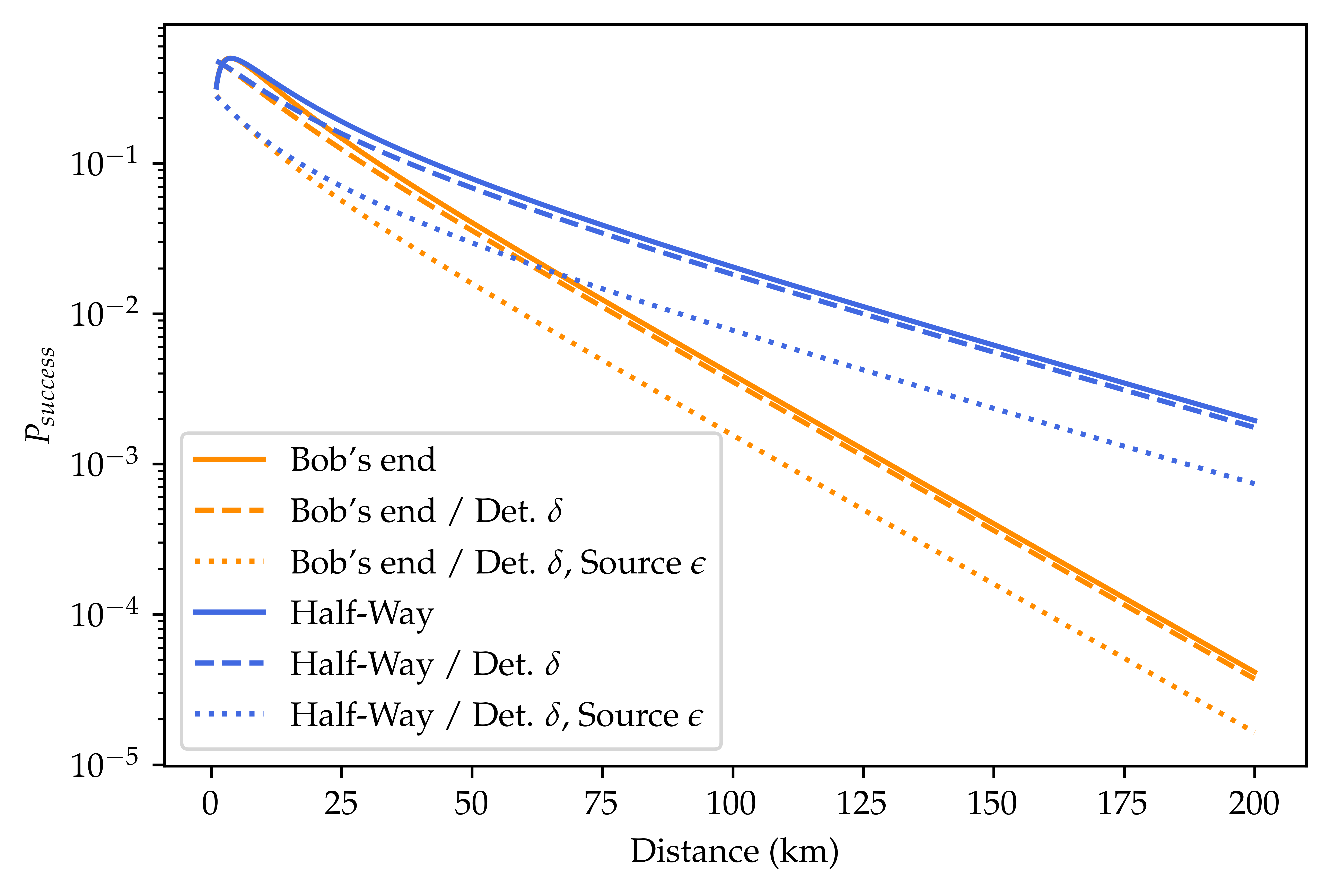}}
\caption{Comparison of the NLA click probabilities in situations \ref{sit1} and \ref{sit3}, using $\delta=0.9, \epsilon=0.9, X=10$. The improvement brought by implementing photon detection half-way between Alice and Bob is entirely maintained in presence of imperfect photon source and detection components. However, with a photon source of quality $\epsilon=0.9$, state purities higher than $X=18$ cannot be reached.}
\label{dist6}
\end{figure}

Treating situation \ref{sit1} first, we find that in the limit of $X \gg 1$ and $\eta \ll 1$, the probabilities of success are respectively 
\begin{equation}
P^{\text{Bob's end}}_{\text{success}} \to \frac{4 \delta \eta}{X}
\end{equation}
 when the measurement is done at Bob's end, and 
 \begin{equation}
 P^{\text{Half-way}}_{\text{success}} \to \frac{2 \delta \sqrt{\eta}}{X}
 \end{equation}
  when it is done half-way between Alice and Bob. The advantage of putting the detectors half-way between Alice and Bob is therefore maintained with the same order of magnitude, as can be seen in Figure \ref{dist6}, and we see that the protocol in general is highly resilient to detection noise. If a photon is lost at the heralding detectors, this simply reduces the success probability with limited effect on output quality. (Exact click probability formulas and other intermediary relationships are given in Appendix \ref{sec:app-nla-imp-detection}.) 

~

Situation \ref{sit3} is unrealistic in the sense that a perfect source would of course be used in both places if one was available. However, our aim is to characterise the limitations introduced by the protocol itself, so looking at its performance with an ideal input is useful. The analysis then reveals additional purification constraints introduced by the protocol even if acting on an initially pure state.  In this case, it is not possible anymore to reach an arbitrarily high state purity, as the imperfection of the source imposes an upper bound on $X$. Independently of whether NLA is implemented at Bob's end or half-way, the bound is always given by
         \begin{equation}
      X_{\text{max}} =   \frac{2   \epsilon}{1-\epsilon},
      \end{equation}

In terms of click probabilities, while a lower source quality certainly causes a significant reduction, the relative scaling of $\sqrt{\eta}$ is preserved in favor of the half-way implementation. (Click probabilities and other intermediary formulas are given in Appendix \ref{sec:app-nla-imp-both}).

\section{Purification Protocol}
\label{sec:purification}

\subsection{Perfect photon source and detection}

\begin{figure}[ht]
\centerline{\includegraphics[width=\columnwidth]{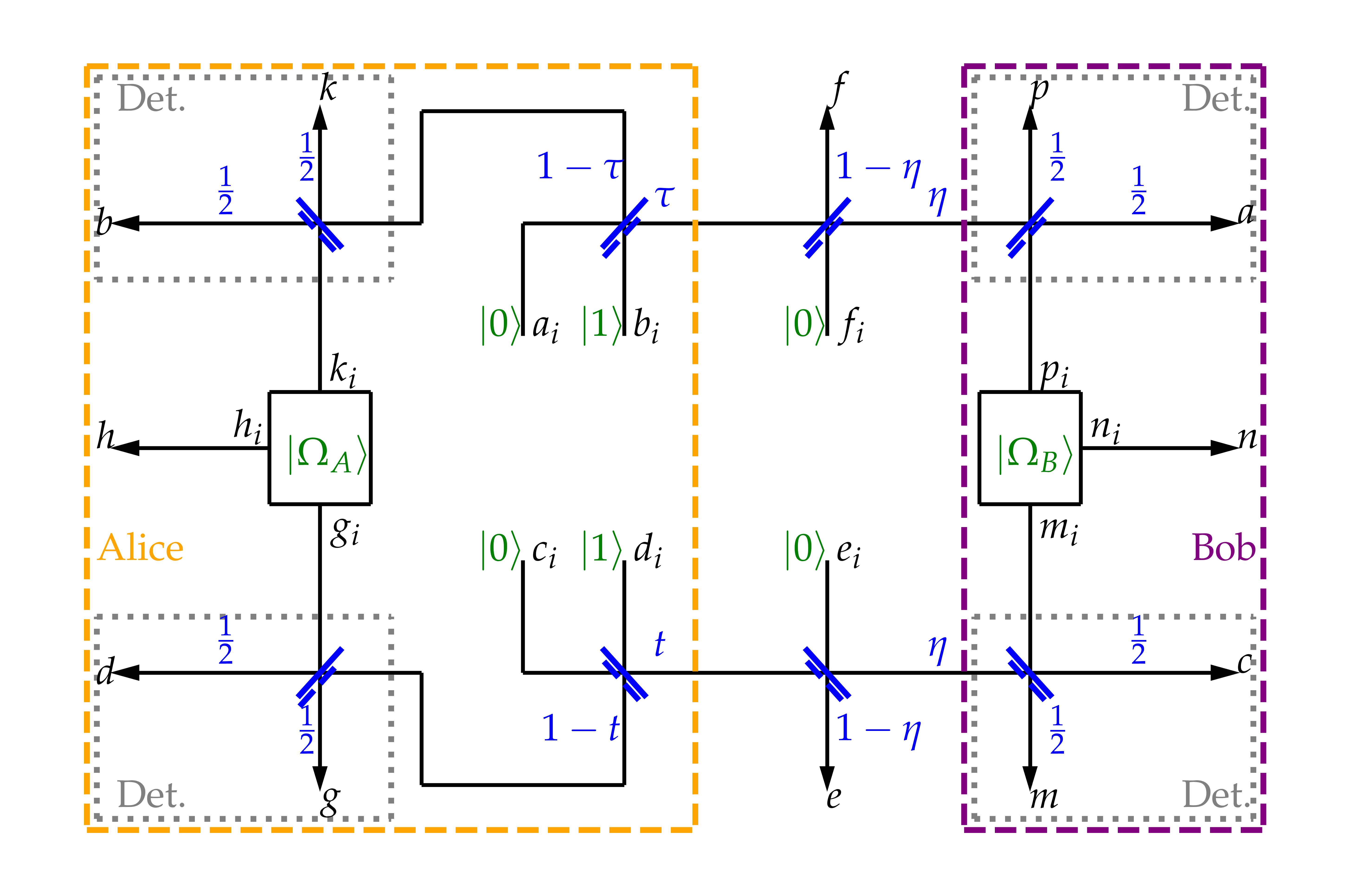}}
\caption{Purification protocol as proposed in \cite{b13}. The resource states $\vert \Omega_A \rangle$ and $\vert \Omega_B \rangle$ are prepared such that a single photon detection at each corner (a ``click'') heralds complete purity of the output state in modes $h$ and $n$. In other words, it is guaranteed that no photon was lost to the environment in modes $e$ and $f$ when a click is obtained.}
\label{diagram_purif}
\end{figure}

We now consider a rather different protocol we will refer to as the purification protocol. The diagram displayed in Figure \ref{diagram_purif} represents the complete purification protocol proposed in \cite{b13}. Key differences to the NLA protocols are that Alice now sends two entangled states through the channel (one gets recovered), and uses more complicated resource states 
  \begin{align}
  \begin{split}
   \vert \Omega_A \rangle &= \sqrt{\frac{1}{2}}  \vert 001 \rangle _{k_i h_i g_i} + \sqrt{\frac{1}{2}}   \vert 110 \rangle _{k_i h_i g_i} \\
 \vert \Omega_B \rangle &=  \sqrt{\frac{1}{2}}  \vert 100 \rangle _{m_i n_i p_i} + \sqrt{\frac{1}{2}}   \vert 011 \rangle _{m_i n_i p_i} 
    \end{split}
   \end{align}

Following the same method as for the NLA protocol, we obtain that an output state projected on no environmental loss gives
    \begin{equation}
  \vert  \psi_f \rangle = - \frac{1}{8} \sqrt{1-t} \sqrt{\tau} \sqrt{\eta} \vert 10 \rangle_{hn} +  \frac{1}{8} \sqrt{\eta} \sqrt{t} \sqrt{1-\tau} \vert 01 \rangle_{hn} 
         \end{equation}
         with probability
     \begin{equation}
   P_f = \frac{1}{64} \eta ( \tau(1-t) + t (1-\tau) )
     \end{equation}
     while the projection on environmental losses yields an empty $\vert \psi_0 \rangle$ and therefore $P_0=0$. 
     As a result, the heralded state is absolutely pure ($X=\infty$). In other words, a click means we know with certainty that no photon was lost to the environment.
     After setting the usual condition of maximum entanglement ($t=\tau$ here) and considering the 16 possible successful states, we obtain
           \begin{equation}
   P_{\text{success}} = \frac{1}{2} \eta t(1-t)
   \end{equation}
   We can set $t=\frac{1}{2}$ to obtain the maximum click probability, and therefore 
          \begin{equation}
          P_{\text{success}}=\frac{\eta}{8}
          \label{purif_psuc_perfect}
   \end{equation}

\subsection{Imperfect photon source and detection}

As in the case of the NLA protocols, the quality of photon sources and detectors must be considered. In line with the earlier assumptions of situation \ref{sit3} in the NLA protocols, we assume that Alice's single-rail entangled states are perfect and only consider the additional noise introduced by the protocol by adding imperfect source (quality $\epsilon$) to the three branches of each resource state $\vert \Omega_A \rangle$ and $\vert \Omega_B \rangle$, and imperfect detection (quality $\delta$) to all 8 photon detectors.
Repeating the earlier analysis, the state purity is now a constant finite quantity given by 
 \begin{equation}
 X = \frac{2 \epsilon^2}{1-\epsilon^2}
 \end{equation}
 and the probability of success is given by
      \begin{equation}
   P_{\text{success}} = \frac{ \delta^4 \epsilon^2 \eta (1+\epsilon^2)}{4 (1+\epsilon)^2}
   \label{purif_psuc_noise}
   \end{equation}
 (Key intermediary results of this derivation are given in Appendix \ref{sec:app-purif}).
 Interestingly, the state purity $X$ produced by this purification protocol can always be reached by the NLA protocols, as $\epsilon \leq1$ guarantees 
 \begin{equation}
 \frac{2 \epsilon^2}{1-\epsilon^2} \leq \frac{ 2 \epsilon}{1-\epsilon}
 \end{equation}

\section{Protocol Evaluation and Comparison}
\label{sec:comparison}

\begin{figure}[ht]
\centerline{\includegraphics[width=\columnwidth]{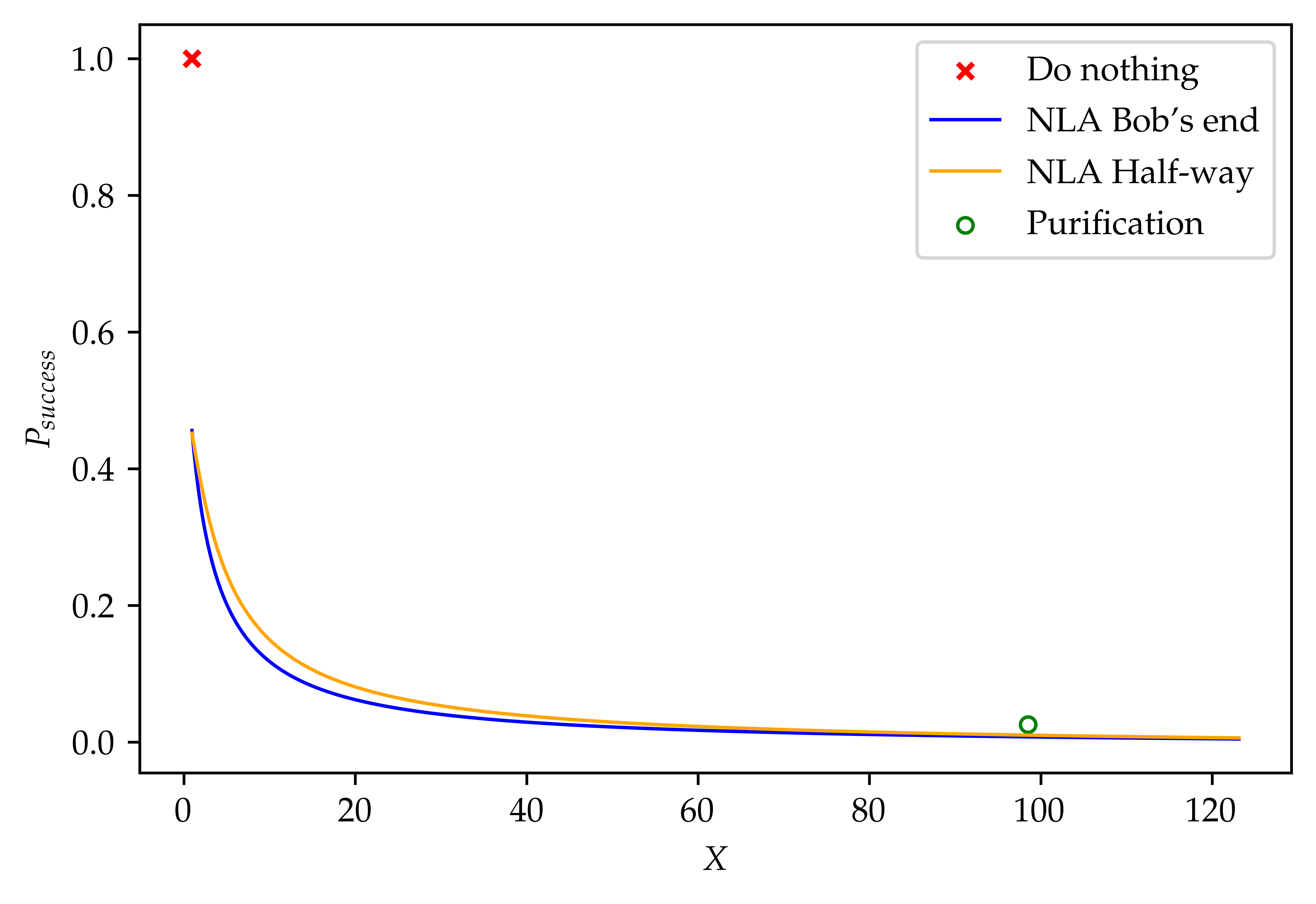}}
\caption{Click probability $P_{\text{success}}$ against state purity $X$, for $d=\SI{25}{\km}$, $\delta=0.9$ and $\epsilon=0.99$. The purification protocol is considered to outperform here, because its metrics place it above the NLA curve trade-off. If the source quality $\epsilon$ is worsened, the green mark moves to the left faster than the NLA curves, such that the half-way NLA soon becomes the best protocol.}
\label{protocol_comp}
\end{figure}

 The properties of the different protocols (or absence thereof) can now be summarized in Table \ref{tab1}, and an example situation is illustrated in Figure \ref{protocol_comp} for $d=\SI{25}{\km}$ and $\epsilon=0.99$.
 
\begin{table}[ht]
\centering
\caption{Limiting behaviour of the different protocols}
\label{table}
\setlength\extrarowheight{5pt}
\begin{tabular}{|M{1.45cm}|M{2.2cm}|M{1.7cm}|M{2.75cm}|}
\hline
Source & Protocol &  Purity $X$ & $P_{\text{success}}$ \\
\hline
\multirow{4}{*}{Perfect} & None &  $\frac{2 \eta}{1-\eta}$ & 1 \\ \cline{2-4}
                                     & NLA Bob's end & Arbitrary & $ \frac{4 \eta}{X} $ \\ \cline{2-4}
                                     & NLA Half-way & Arbitrary & $ \frac{2 \sqrt{\eta}}{X} $ \\ \cline{2-4}
                                     & Purification & $\infty$ & $ \frac{\eta}{8} $  \\
\hline
\multirow{4}{*}{Quality $\epsilon$} & None &  $\frac{2 \eta}{1-\eta}$ & 1 \\ \cline{2-4}  
                                     & NLA Bob's end & Up to $\frac{2 \epsilon}{1-\epsilon}$ & $4\delta \eta  \left( \frac{1}{X} - \frac{1-\epsilon}{2} \right)$  \\ \cline{2-4} 
                                     & NLA Half-way & Up to $\frac{2 \epsilon}{1-\epsilon}$ & $2\delta \sqrt{\eta} \left( \frac{1}{X} - \frac{1-\epsilon}{2} \right) $ \\ \cline{2-4} 
                                     & Purification & $\frac{2 \epsilon^2}{1-\epsilon^2}$ & $ \frac{\delta^4 \epsilon^2 \eta }{8} $ \\
\hline
\multicolumn{4}{p{\columnwidth}}{The limiting behaviour for the click probability $P_{\text{success}}$ is obtained under the assumptions $X \gg 1$, $\eta \ll 1$  and $1 - \epsilon \ll 1$. The$\sqrt{\eta}$ scaling in $P_{\text{success}}$ is critical to the outperformance of half-way NLA vs the other 2 protocols, and ultimately dominates any other effect at long distances, unless the detection and source quality is absolutely perfect.}
\end{tabular}
\label{tab1}
\end{table}

Firstly, it is clear that if $\epsilon < \eta$, doing nothing produces a purer state and a higher $P_{\text{success}}$ than any of the protocols considered, and is therefore a better choice. The protocols would simply be introducing more noise in the system than there originally was. The scaling of $\sqrt{\eta}$ for NLA half-way vs $\eta$ for the other two protocols also means that, whatever the other parameters might be, there is always going to be some distance far enough such that this protocol will outperform the others. It also looks likely that the purification protocol may only outperform with very good source quality, since both $X$ and $P_{\text{success}}$ are scaling with $\epsilon^2$, instead of $\epsilon$ in NLA. 

~

In order to produce a consistent comparison of the three protocols accross distances and source quality, a numerical analysis is then performed using the following logic: 
\begin{itemize}[noitemsep]
\item if $\epsilon < \eta$, doing nothing is best.
\item otherwise, for a given source quality $\epsilon$, the constant purity ratio $X$ obtained in the purification protocol is calculated, and fixed at the same level for the two NLA protocols
\item the success probability is calculated as a function of distance for all three protocols, and a higher success probability for a given distance is considered to be a superior protocol.
\end{itemize}

~

\begin{figure}[hbt!]
\centerline{\includegraphics[width=\columnwidth]{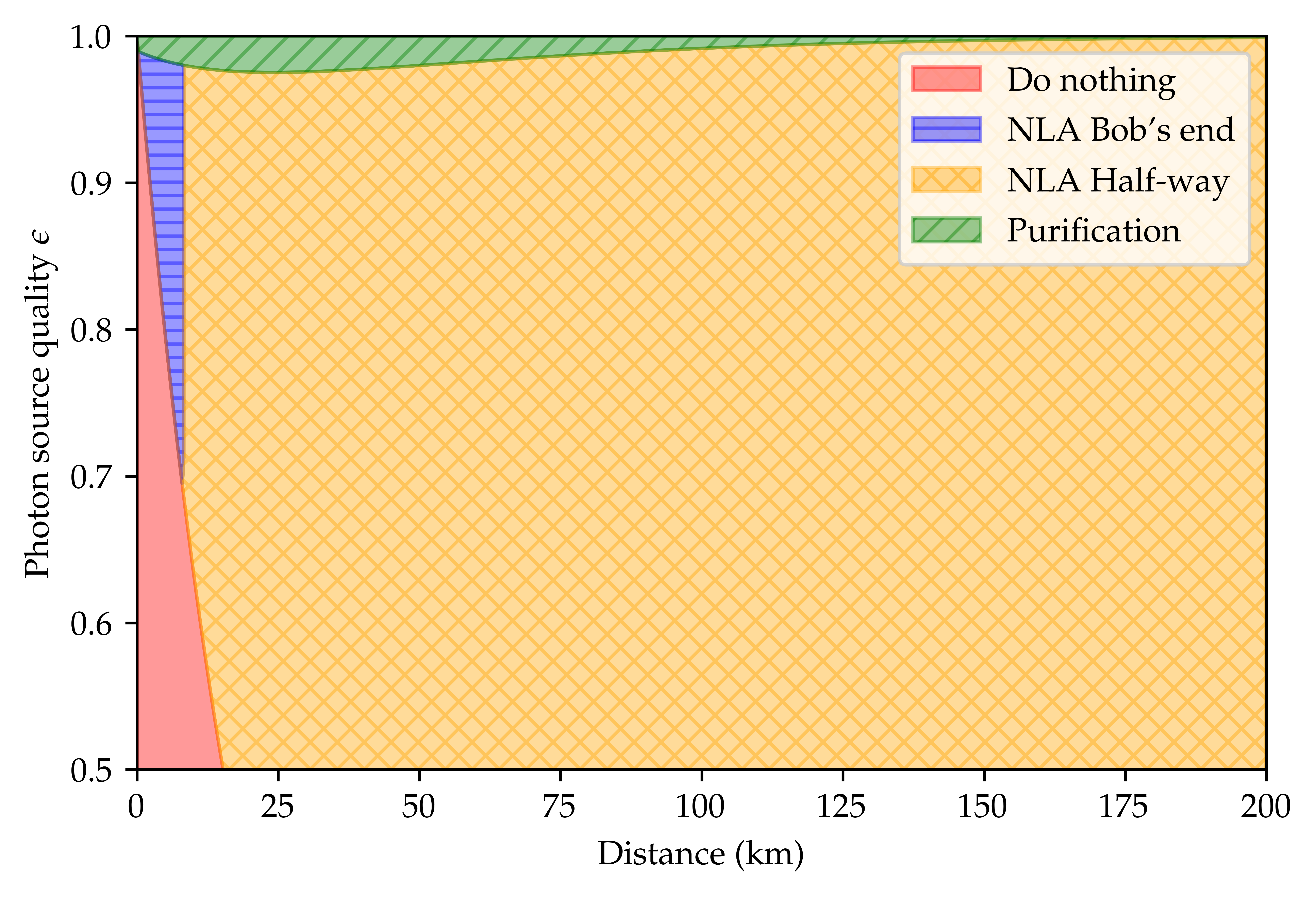}}
\caption{Best-performing protocol accross distances and source quality (detection quality $\delta$ is set to $0.9$). The half-way NLA protocol offers the best performance in most situations. Exceptions are for very low distances, where doing nothing or NLA at Bob's end may be better, or for very high-quality photon source components, where the purification protocol outperforms up to a certain distance.}
\label{areabound}
\end{figure}

Results of the analysis are displayed in Figure \ref{areabound} for $\delta=0.9$. The half-way NLA protocol appears to be superior in most situations, as the distance scaling dominates. Even in the situations where one of the other protocols is superior, it typically only appears to offer marginal improvements against the half-way NLA protocol. An exception to this rule might be with the advent of future precise source technologies with $\epsilon \approx 1$, whereby operating in the "thin-green-slice" of Figure \ref{areabound} may result in a significant advantage.

~
~

\section{Conclusion}
\label{sec:conclusion}

We compared the theoretical performance of three quantum communication protocols in terms of output state purity and probability of success, under different assumptions of photon source and detection quality. In most realistic situations, the NLA protocol implemented in a central station between Alice and Bob offers better trade-off metrics than NLA implemented at Bob's or the purification protocol. This is primarily due to its probability of success scaling with the square root of the channel transmissivity, while the other two protocols are linear. The purification protocol offers the promise of perfect output state purity, but it is less resilient to source noise than the other protocols due to the higher number of components involved. Building effective quantum communication protocols is an area of active current research, and we expect our results will be of interest in providing a quantitative comparison framework of different schemes. In future work, the analysis may be extended to take into account other experimental considerations such as dark counts and thermal noise, as well as to cover a broader set of protocols.

~
\pagebreak

\onecolumngrid

\appendix

\section{NLA with perfect photon source and detection}
\label{app:nla-perfect}

\subsection{NLA at Bob's end}
\label{app:nla-perfect-ref}

We provide full calculations for this first setup where the detections are done at Bob's end and perfect photon source and detection components are available. Intermediary results for the other setups are given without the detailed derivation but follow exactly the same logic.

The input state is 
\begin{equation}
 \vert 01010 \rangle _{a_i b_i c_i d_i e_i } = b_i ^\dagger d_i ^\dagger \vert 00000 \rangle  _{a_i b_i c_i d_i e_i }
 \end{equation}
 The input operators $b_i ^\dagger$ and  $d_i ^\dagger$ can be written in terms of the output operators based on the diagram from Figure \ref{diag1}:
 \begin{align}
 b_i ^\dagger &= 
 \sqrt{\tau} a^\dagger 
 + \sqrt{1-\tau} \sqrt{1-\eta} e^\dagger 
 +   \sqrt{1-\tau}\sqrt{\eta} \sqrt{\frac{1}{2}} c^\dagger 
 + \sqrt{1-\tau} \sqrt{\eta} \sqrt{\frac{1}{2}} b^\dagger \\
 d_i^\dagger &= 
 \sqrt{t} d^\dagger
  - \sqrt{1-t} \sqrt{\frac{1}{2}} b^\dagger 
 + \sqrt{1-t}\sqrt{\frac{1}{2}} c^\dagger
 \end{align}
 Now we can write the input state in terms of the output operators
\begin{align}
\begin{split}
  \vert 01010 \rangle _{a_i b_i c_i d_i e_i } = &\biggl(
   \sqrt{\tau}  \sqrt{t}  a^\dagger d^\dagger
  - \sqrt{\tau} \sqrt{1-t} \sqrt{\frac{1}{2}} a^\dagger b^\dagger 
  +  \sqrt{\tau}  \sqrt{1-t}\sqrt{\frac{1}{2}} a^\dagger c^\dagger \\
 &+ \sqrt{1-\tau} \sqrt{1-\eta}   \sqrt{t} d^\dagger e^\dagger 
 - \sqrt{1-\tau} \sqrt{1-\eta} \sqrt{1-t} \sqrt{\frac{1}{2}} b^\dagger  e^\dagger  
 + \sqrt{1-\tau} \sqrt{1-\eta}   \sqrt{1-t}\sqrt{\frac{1}{2}} c^\dagger e^\dagger \\ 
  &+   \sqrt{1-\tau}\sqrt{\eta} \sqrt{\frac{1}{2}}   \sqrt{t} c^\dagger  d^\dagger
   -   \sqrt{1-\tau}\sqrt{\eta} \sqrt{\frac{1}{2}} \sqrt{1-t} \sqrt{\frac{1}{2}} c^\dagger   b^\dagger 
    +   \sqrt{1-\tau}\sqrt{\eta} \sqrt{\frac{1}{2}}  \sqrt{1-t}\sqrt{\frac{1}{2}} \left(c^\dagger\right)^2 \\
 &+ \sqrt{1-\tau} \sqrt{\eta} \sqrt{\frac{1}{2}} \sqrt{t}  b^\dagger  d^\dagger
  - \sqrt{1-\tau} \sqrt{\eta} \sqrt{\frac{1}{2}}  \sqrt{1-t} \sqrt{\frac{1}{2}} \left( b^\dagger \right)^2
   + \sqrt{1-\tau} \sqrt{\eta} \sqrt{\frac{1}{2}}  \sqrt{1-t}\sqrt{\frac{1}{2}} b^\dagger  c^\dagger \biggr) \\
  &\vert 00000 \rangle  _{a_i b_i c_i d_i e_i }   
  \end{split}
\end{align}
After applying the operators to the state, we get
\begin{align}
\begin{split}
  \vert 01010 \rangle _{a_i b_i c_i d_i e_i } = &
      \sqrt{\tau}  \sqrt{t} \vert 10010 \rangle  _{abcde}
  - \sqrt{\tau} \sqrt{1-t} \sqrt{\frac{1}{2}} \vert 11000 \rangle  _{abcde}
  +  \sqrt{\tau}  \sqrt{1-t}\sqrt{\frac{1}{2}} \vert 10100 \rangle  _{abcde} \\
 &+ \sqrt{1-\tau} \sqrt{1-\eta}   \sqrt{t} \vert 00011 \rangle  _{abcde}
 - \sqrt{1-\tau} \sqrt{1-\eta} \sqrt{1-t} \sqrt{\frac{1}{2}} \vert 01001 \rangle  _{abcde}  \\
 &+ \sqrt{1-\tau} \sqrt{1-\eta}   \sqrt{1-t}\sqrt{\frac{1}{2}} \vert 00101 \rangle  _{abcde}  
  +   \sqrt{1-\tau}\sqrt{\eta} \sqrt{\frac{1}{2}}   \sqrt{t} \vert 00110 \rangle  _{abcde} \\
   &-   \sqrt{1-\tau}\sqrt{\eta} \sqrt{\frac{1}{2}} \sqrt{1-t} \sqrt{\frac{1}{2}} \vert 01100 \rangle  _{abcde}
    +   \sqrt{1-\tau}\sqrt{\eta} \sqrt{\frac{1}{2}}  \sqrt{1-t}\sqrt{\frac{1}{2}} \sqrt{2} \vert 00200 \rangle  _{abcde} \\
 &+ \sqrt{1-\tau} \sqrt{\eta} \sqrt{\frac{1}{2}} \sqrt{t}  \vert 01010 \rangle  _{abcde}
  - \sqrt{1-\tau} \sqrt{\eta} \sqrt{\frac{1}{2}}  \sqrt{1-t} \sqrt{\frac{1}{2}} \sqrt{2} \vert 02000 \rangle  _{abcde} \\
   &+ \sqrt{1-\tau} \sqrt{\eta} \sqrt{\frac{1}{2}}  \sqrt{1-t}\sqrt{\frac{1}{2}} \vert 01100 \rangle  _{abcde} 
  \end{split}
\end{align}
  If we project onto $_b \langle 0 \vert$ and $_c \langle 1 \vert$ (ie, one of the two click states), most terms drop out and we get the state
   \begin{equation}
  \vert \psi \rangle =  \sqrt{\tau}  \sqrt{1-t}\sqrt{\frac{1}{2}} \vert 100 \rangle  _{ade}
+ \sqrt{1-\tau} \sqrt{1-\eta}   \sqrt{1-t}\sqrt{\frac{1}{2}} \vert 001 \rangle  _{ade}  
  +   \sqrt{1-\tau}\sqrt{\eta} \sqrt{\frac{1}{2}}   \sqrt{t} \vert 010 \rangle  _{ade} 
    \end{equation}
    If instead we project onto ${}_b \langle 1 \vert$ and ${}_c \langle 0 \vert$ (the other successful BSM state), we get
     \begin{equation}
  \vert \psi_{2} \rangle = - \sqrt{\tau}  \sqrt{1-t}\sqrt{\frac{1}{2}} \vert 100 \rangle  _{ade}
- \sqrt{1-\tau} \sqrt{1-\eta}   \sqrt{1-t}\sqrt{\frac{1}{2}} \vert 001 \rangle  _{ade}  
  +   \sqrt{1-\tau}\sqrt{\eta} \sqrt{\frac{1}{2}}   \sqrt{t} \vert 010 \rangle  _{ade} 
    \end{equation}
    The probabilities of both states are the same since the amplitudes only differ by an experimentally correctable sign change, so the overall probability of success is
\begin{equation}
P_{\text{success}} = 2 \langle \psi \vert \psi \rangle = \tau \left(1-t \right) + \left( 1- \tau \right) \left( 1 - \eta \right)  \left( 1 - t \right)  +\left( 1 - \tau \right) \eta  t 
\label{psuccess_orig_nonoise}
 \end{equation}
 If we want to express the density operator for $\vert \psi \rangle$ as a mixed state conditional on environment outcome, we can see that projecting $\vert \psi \rangle$ onto $_e \langle 0 \vert$ gives
     \begin{equation}
      \vert \psi_{f} \rangle  = \sqrt{\tau}  \sqrt{1-t}\sqrt{\frac{1}{2}} \vert 10 \rangle  _{ad}
  +   \sqrt{1-\tau}\sqrt{\eta} \sqrt{\frac{1}{2}}   \sqrt{t} \vert 01 \rangle  _{ad} 
         \end{equation}
         with probability
     \begin{equation}
 P_f = \langle \psi_{f} \vert \psi_{f} \rangle     =  \frac{1}{2}  \tau \left(1-t \right) +  \frac{1}{2}  \left( 1-\tau \right) \eta t
     \end{equation}
and projecting $\vert \psi \rangle$ onto $_e \langle 1 \vert$ gives
 \begin{equation}
 \vert \psi_{0} \rangle =
\sqrt{1-\tau} \sqrt{1-\eta}   \sqrt{1-t} \sqrt{\frac{1}{2}} \vert 00 \rangle  _{ad}  
\label{psizero_orig_nonoise}
\end{equation}
with probability
     \begin{equation}
 P_0 =  \langle \psi_{0} \vert \psi_{0} \rangle     =   \frac{1}{2} \left(1-\tau \right)  \left(1-\eta \right)   \left(1-t \right) 
 \end{equation}    
As expected, we have
 \begin{equation}
     P_{\text{success}}  = 2 \left ( P_0 +  P_f \right)
     \end{equation}
The normalized density operator can be written in the form:
     \begin{equation}
  \hat{\rho} = \frac{ P_0    \vert    \psi_{0}^{(N)}   \rangle  \langle  \psi_{0}^{(N)} \vert  + P_f    \vert    \psi_{f}^{(N)}  \rangle  \langle   \psi_{f}^{(N)} \vert } {P_0 + P_f} 
  \end{equation}
  where $ \vert    \psi_{0}^{(N)}   \rangle$ and  $\vert    \psi_{f}^{(N)}   \rangle$ are the normalized states, such that
       \begin{equation}
   \vert    \psi_{0}^{(N)}   \rangle =  \vert 00 \rangle  _{ad}
   \end{equation}
   and
        \begin{equation}
        \vert    \psi_{f}^{(N)}  \rangle = \frac{\sqrt{\tau}  \sqrt{1-t}\sqrt{\frac{1}{2}} \vert 10 \rangle  _{ad}
  +   \sqrt{1-\tau}\sqrt{\eta} \sqrt{\frac{1}{2}}   \sqrt{t} \vert 01 \rangle  _{ad}  }{\sqrt{\tau}  \sqrt{1-t}\sqrt{\frac{1}{2}} +
   \sqrt{1-\tau}\sqrt{\eta} \sqrt{\frac{1}{2}}}
    \end{equation}         
If we want   $\vert \psi_{f} \rangle$ to be in the maximally entangled state, we need to set $\tau$ such that
 \begin{equation}
\sqrt{\tau} \sqrt{1-t} = \sqrt{1-\tau} \sqrt{\eta} \sqrt{t}
 \end{equation} 
 so 
  \begin{equation}
  \tau = \frac{t \eta}{1 - t + t \eta}
  \end{equation}
Now we can replace $\tau$ in Equation \ref{psuccess_orig_nonoise}, and we get
    \begin{equation}
    P_{\text{success}} = (1-t) (1 - \eta + \frac{t \eta^2}{1-t +t \eta} ) + (1 - \frac{ t \eta}{1-t + t \eta} ) \eta t
    \label{interm_orig_nonoise}
   \end{equation} 
From the expressions for $P_0$ and $P_f$, we can also write $X = \frac{P_f}{P_0}$ as a function of $t$ and $\eta$ and after some simplifications, we get
    \begin{equation}
       X  = \frac{ 2 t \eta}{(1-\eta)(1-t)}
    \label{x_eq_orig_nonoise}
          \end{equation} 
Inverting this to get $t$ as a function of $X$ and $\eta$, we obtain
\begin{equation}
t = \frac{X ( 1 - \eta) }{2 \eta + X ( 1- \eta) }
\end{equation}
Now we can replace all occurrences of $t$ by their expressions in Equation \ref{interm_orig_nonoise}, and after some manipulations, we obtain
\begin{equation}
\boxed{
 P_{\text{success}} = \frac{4 \eta  (1 - \eta) (1+X) } { \left( 2 \eta + X ( 1- \eta)  \right) \left( 2 + X ( 1- \eta)  \right)}
 }
\end{equation}
For $\eta \ll 1$ and $X \gg 1$, this expression reduces to
 \begin{equation}
  P_{\text{success}} \approx \frac{4 \eta }{X}
   \end{equation}
We note the maximum value of $P_{\text{success}}$ is reached for an $X$ that satisfies 
\begin{equation}
\frac{ \partial  P_{\text{success}}} {\partial X}=0
\end{equation}
 which gives
 \begin{equation}
\left( 2 \eta + X ( 1- \eta)  \right) \left( 2 + X ( 1- \eta)  \right)   -  (1-\eta) (1+X) \left(  2 +   2 \eta  + 2X ( 1- \eta)  \right) =0
 \end{equation}
This is a polynomial of degree 2 in $X$, and after collecting the terms and solving for the roots, we get
 \begin{equation}
X_{\text{root}} = -1 \pm \frac{\sqrt{ (\eta + 1) ( 3 \eta -1)}} {1 - \eta}
\label{xmax_orig_nonoise}
    \end{equation}
    From using $\eta \leq 1$, we get $3 \eta - 1 \leq \eta+1$ and therefore 
     \begin{equation}
    X_{\text{root}} \leq \frac{2 \eta}{1-\eta},
        \end{equation}
     showing that any $X$ obtained in the left portion of the curve in Figure \ref{psuc1} (where the probability of success increases with increasing $X$) is actually always smaller than the $X$ obtained from doing nothing.

\subsection{NLA half-way}
\label{app:nla-perfect-alt}

Following the same logic as in subsection \ref{app:nla-perfect-ref}, we obtain the following key intermediary results.
The click probability is given by
        \begin{equation}
P_{\text{success}} =   
 \tau \sqrt{\eta} \left(1-t \right) + 
  \left( 1- \tau \right) \left( 1 - \sqrt{\eta} \right)  \left( 1 - t \right) \sqrt{\eta} 
+   \left( 1 - \tau \right) \sqrt{\eta} t 
+ \left( 1 - \tau \right) \sqrt{\eta} (1-t) (1 - \sqrt{\eta})    
\label{psuccess_mid_nonoise}
        \end{equation}        
        The projection of a click state $\vert \psi \rangle$ onto no environmental losses gives
           \begin{equation}
      \vert \psi_{f} \rangle  = \sqrt{\tau}  \sqrt{\sqrt{\eta}} \sqrt{1-t}\sqrt{\frac{1}{2}} \vert 10 \rangle  _{ad}
  +   \sqrt{1-\tau} \sqrt{\sqrt{\eta}} \sqrt{\frac{1}{2}}   \sqrt{t} \vert 01 \rangle  _{ad} 
         \end{equation}
         with probability
     \begin{equation}
 P_f = \langle \psi_{f} \vert \psi_{f} \rangle     =  \frac{1}{2}  \tau \sqrt{\eta} \left(1-t \right) +  \frac{1}{2}  \left( 1-\tau \right) \sqrt{\eta} t
     \end{equation}
  while the probability of obtaining a state that includes environmental losses after a click is
     \begin{equation}
P_0 =   \left(1-\tau \right)  \left(1- \sqrt{\eta} \right)   \left(1-t \right)  \sqrt{\eta}
 \end{equation}
 The condition of a maximally entangled $ \vert \psi_{f} \rangle $ yields 
  \begin{equation}
 \tau = t
   \end{equation}
   With this condition, the click probability is given by
       \begin{align}
    P_{\text{success}}  = 2 t \sqrt{\eta} (1-t) + 2 (1-t)^2 \sqrt{\eta} (1-\sqrt{\eta}) 
    \label{interm_mid_nonoise}
   \end{align} 
The ratio $X$ representing state purity is
     \begin{equation}
        X = \frac{ t }{(1-t) (1 - \sqrt{\eta}) }
         \end{equation}
and this can be inverted to obtain $t$ as a function of $X$, such that
\begin{equation}
t = \frac{X (\sqrt{\eta} - 1) }{  X (\sqrt{\eta} - 1) -1 }
\end{equation}
After replacing all occurrences of $t$ by their expressions in Equation \ref{interm_mid_nonoise} and some manipulations, we get
  \begin{equation}
\boxed{
    P_{\text{success}}   = \frac{  2 \sqrt{\eta} (1-\sqrt{\eta}) (1+X)}{  (X (\sqrt{\eta} - 1) -1)^2  } 
   }
   \label{psuccess_mid_nonoise}
 \end{equation}
 For $X \gg 1$ and $\eta \ll 1$, this expression reduces to
\begin{equation}
 P_{\text{success}} \approx \frac{2 \sqrt{\eta} }{X}
 \end{equation}

\section{NLA with imperfect photon detection}
\label{sec:app-nla-imp-detection}

\begin{figure}[h]
\subfloat[NLA at Bob's end \label{diag3}]{\includegraphics[width = 0.5 \columnwidth]{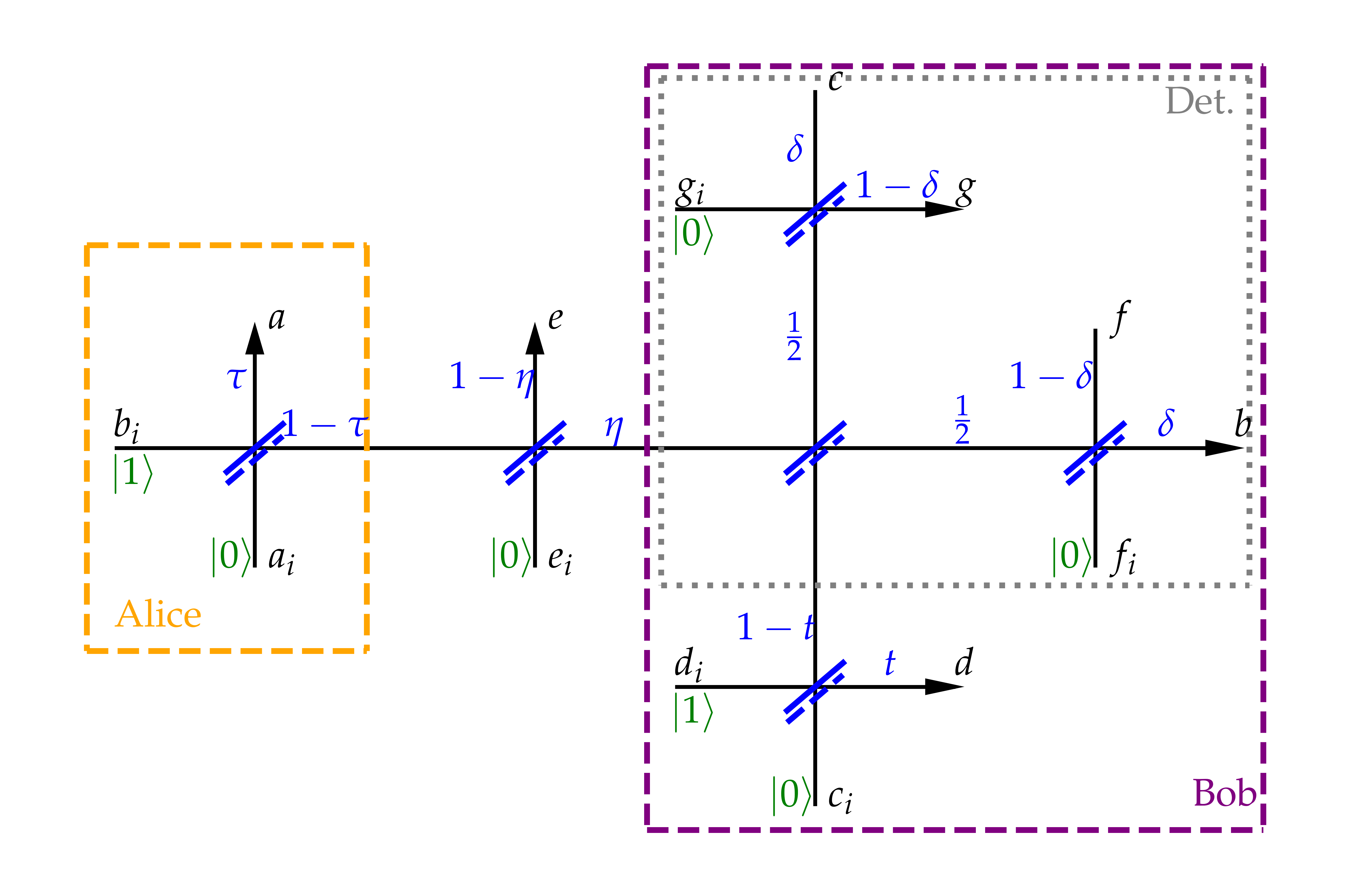}} 
\subfloat[NLA half-way \label{diag4}]{\includegraphics[width = 0.5 \columnwidth]{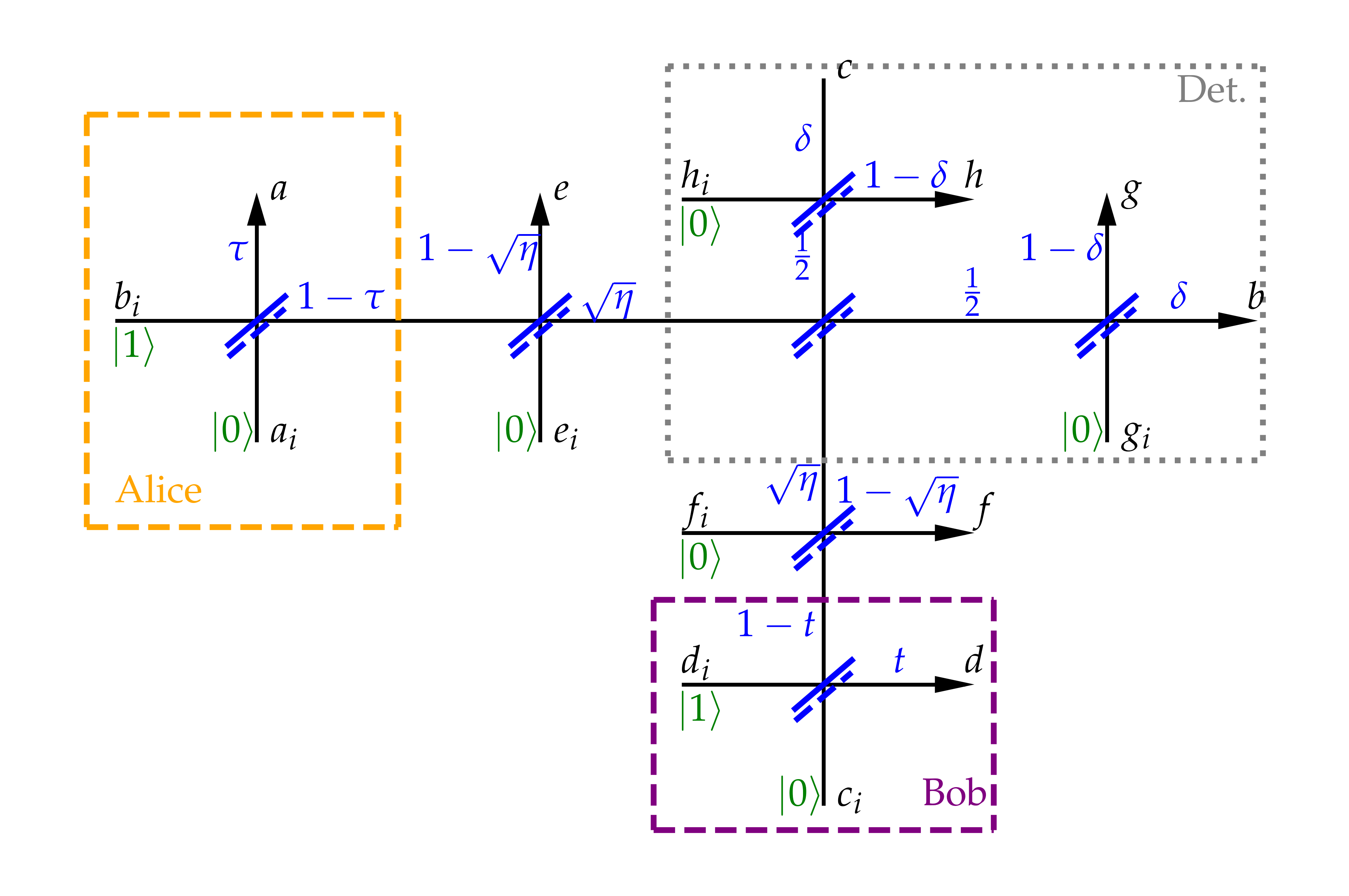}}
\caption{NLA protocol in presence of imperfect detection components. Detection noise is modelled as an additional vacuum state entering the open port of a beamsplitter of reflection $\delta$, such that a photon going through this component is detected with probability $\delta$ and lost with probability $1-\delta$.}
\end{figure}

\subsection{NLA at Bob's end}

Using Figure \ref{diag3} and the same logic as in Appendix \ref{app:nla-perfect}, we obtain the following key intermediary results. The click probability is given by
        \begin{equation}
P_{\text{success}}    = \delta \tau  \left(1-t \right) + \delta \left( 1- \tau \right) \left( 1 - \eta \right)  \left( 1 - t \right)  + \delta \left( 1 - \tau \right) \eta  t  + 2 ( 1- \tau) (1-t) (1- \delta) \eta \delta
\label{psuccess_end_meas_noise}
            \end{equation}
The projection of a click state $\vert \psi \rangle$ onto no environmental losses gives
     \begin{equation}
 P_f = \langle \psi_{f} \vert \psi_{f} \rangle     =  \frac{1}{2}  \tau \delta \left(1-t \right) +  \frac{1}{2}  \left( 1-\tau \right) \delta \eta t
     \end{equation}
        while the probability of obtaining a state that includes environmental losses after a click is
           \begin{equation}
P_0 =  \delta (1-\tau) (1-t) \left( \frac{1}{2} (1-\eta) + \eta ( 1-\delta) \right)
\end{equation}
The condition of a maximally entangled $ \vert \psi_{f} \rangle $ is unchanged from Section \ref{app:nla-perfect-ref}
          \begin{equation}
  \tau = \frac{t \eta}{1 - t + t \eta}
  \end{equation}
With this condition, the click probability is given by
     \begin{equation}
    P_{\text{success}} = \delta (1-t) (1 - \eta + \frac{t \eta^2}{1-t +t \eta} ) + \delta (1 - \frac{ t \eta}{1-t + t \eta} ) \eta t
    + 2 (1-t) (1-\delta) \eta \delta \left( \frac{ 1-t}{1-t +t \eta} \right)
    \label{interm_orig_meas_noise}
   \end{equation} 
 The ratio $X$ representing state purity is
      \begin{equation}
       X  = \frac{ 2 t \eta}{(1+\eta - 2 \eta \delta)(1-t)}
    \label{x_eq_orig_meas_noise}
          \end{equation} 
and this can be inverted to obtain $t$ as a function of $X$, such that
 \begin{equation}
t = \frac{X ( 1 + \eta - 2 \eta \delta) }{2 \eta + X ( 1 + \eta - 2 \eta \delta) }
\end{equation}
After replacing all occurrences of $t$ by their expressions in Equation \ref{interm_orig_meas_noise} and some simplifications, we get
   \begin{equation}
\boxed{
    P_{\text{success}}   = \frac{4  \delta \eta (1 + \eta - 2 \eta \delta) (1+X) } { \left( 2 \eta + X ( 1 + \eta - 2 \eta \delta)  \right) \left( 2 + X ( 1 +  \eta - 2 \eta \delta)  \right) }
   }
   \label{psuccess_final_orig_meas_noise}
 \end{equation}
For $\eta \ll 1$ and $X \gg 1$, this expression reduces to
 \begin{equation}
 P_{\text{success}} \approx \frac{4 \delta \eta }{X}
 \end{equation}
  
\subsection{NLA half-way}

For this case as represented in Figure \ref{diag4}, the click probability is 
  \begin{equation}
P_{\text{success}}    = \delta \sqrt{\eta} \left( (1- \tau) t + \tau ( 1-t) \right) + 2 \delta \sqrt{\eta} (1-\tau)(1-t) \left(1- \delta \sqrt{\eta} \right)
    \label{psuccess_mid_meas_noise}
 \end{equation}
The projection of a click state $\vert \psi \rangle$ onto no environmental losses gives
         \begin{equation}
 P_f = \frac{1}{2} \sqrt{\eta} \delta \left( (1-\tau) t + \tau ( 1-t) \right)
 \end{equation}
 and projecting onto environmental losses gives
    \begin{equation}
P_0 =  (1-\tau)(1-t)\delta \sqrt{\eta} (1-\delta \sqrt{\eta} )
\end{equation}
 The state $  \vert \psi_{f} \rangle $ can be written in the same way as in section \ref{app:nla-perfect-alt}, with only a factor of $\sqrt{\delta}$ in front of it. The condition on $\tau$ to reach maximum entanglement is therefore unchanged
\begin{equation}
 \tau = t
   \end{equation}
With this condition, the click probability is given by
      \begin{equation}
P_{\text{success}}    = 2 \delta \sqrt{\eta} t (1-t) +  2 \delta \sqrt{\eta} (1-t)^2 \left(1- \delta \sqrt{\eta}  \right)
    \label{interm_mid_meas_noise}
 \end{equation}
 The ratio $X$ representing state purity is given by
      \begin{equation}
       X  = \frac{t}{(1-t) (1-\delta \sqrt{\eta} )}
    \label{x_eq_mid_meas_noise}
          \end{equation} 
 and this can be inverted to obtain $t$ as a function of $X$, such that
  \begin{equation}
t = \frac{X (1-\delta \sqrt{\eta} ) }{1 + X (1-\delta \sqrt{\eta}) }
\end{equation}
After replacing all occurrences of $t$ by their expressions in Equation \ref{interm_mid_meas_noise} and some simplifications, we get
 \begin{equation}
\boxed{
    P_{\text{success}}   = \frac{2 \delta \sqrt{\eta} (1+X) (1-\delta \sqrt{\eta})}
    { \left( 1 + X (1-\delta \sqrt{\eta} )  \right)^2}
    }
   \label{psuccess_final_mid_meas_noise}
 \end{equation}
  For $\eta \ll 1$ and $X \gg 1$, this expression reduces to
\begin{equation}
 P_{\text{success}} \approx \frac{2 \delta \sqrt{\eta} }{X}
 \end{equation}

\section{Imperfect photon source at Bob's end and imperfect photon detection}
\label{sec:app-nla-imp-both}

\begin{figure}[h]
\subfloat[NLA at Bob's end \label{diag5}]{\includegraphics[width = 0.5 \columnwidth]{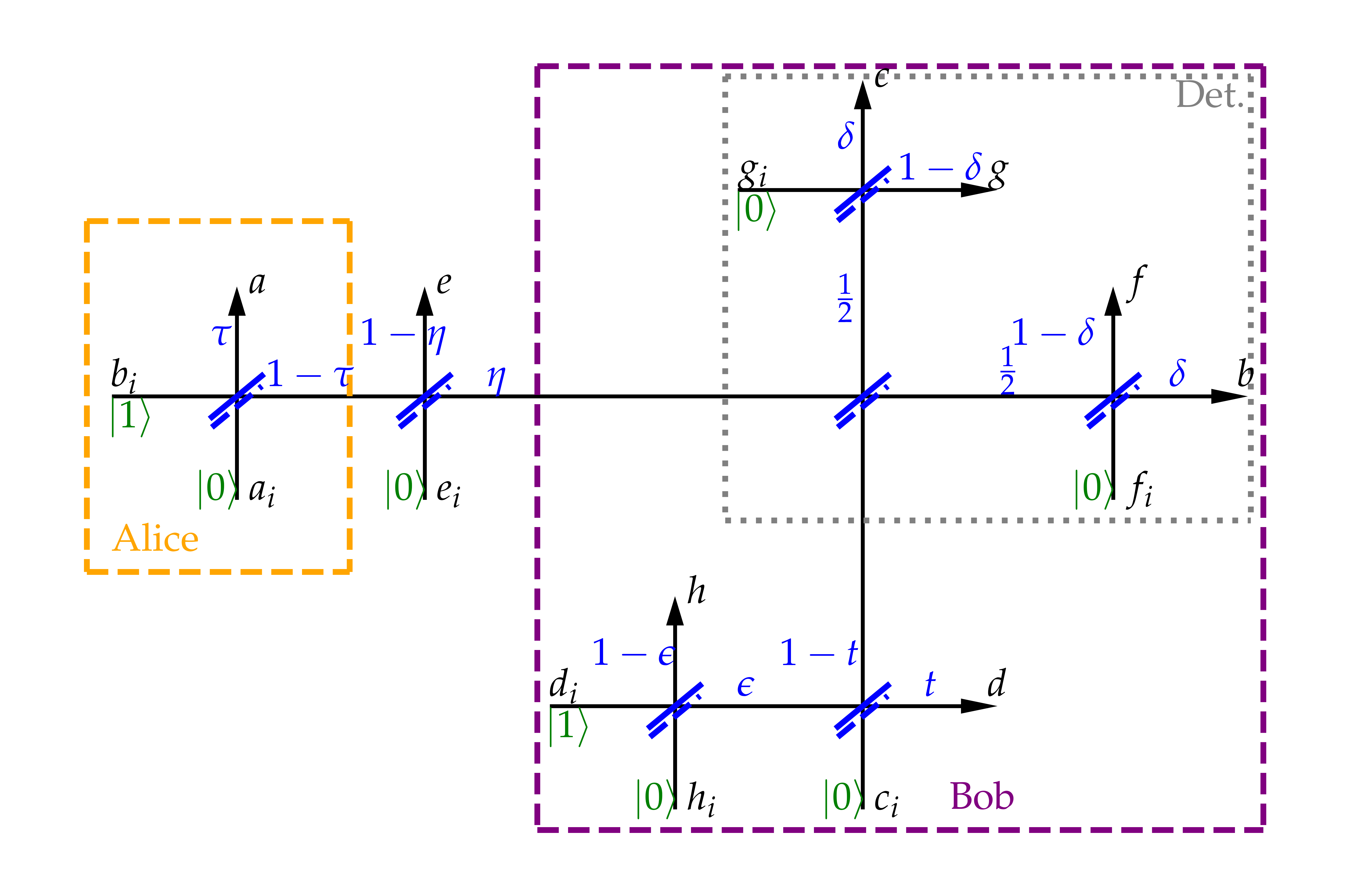}} 
\subfloat[NLA half-way \label{diag6}]{\includegraphics[width = 0.5 \columnwidth]{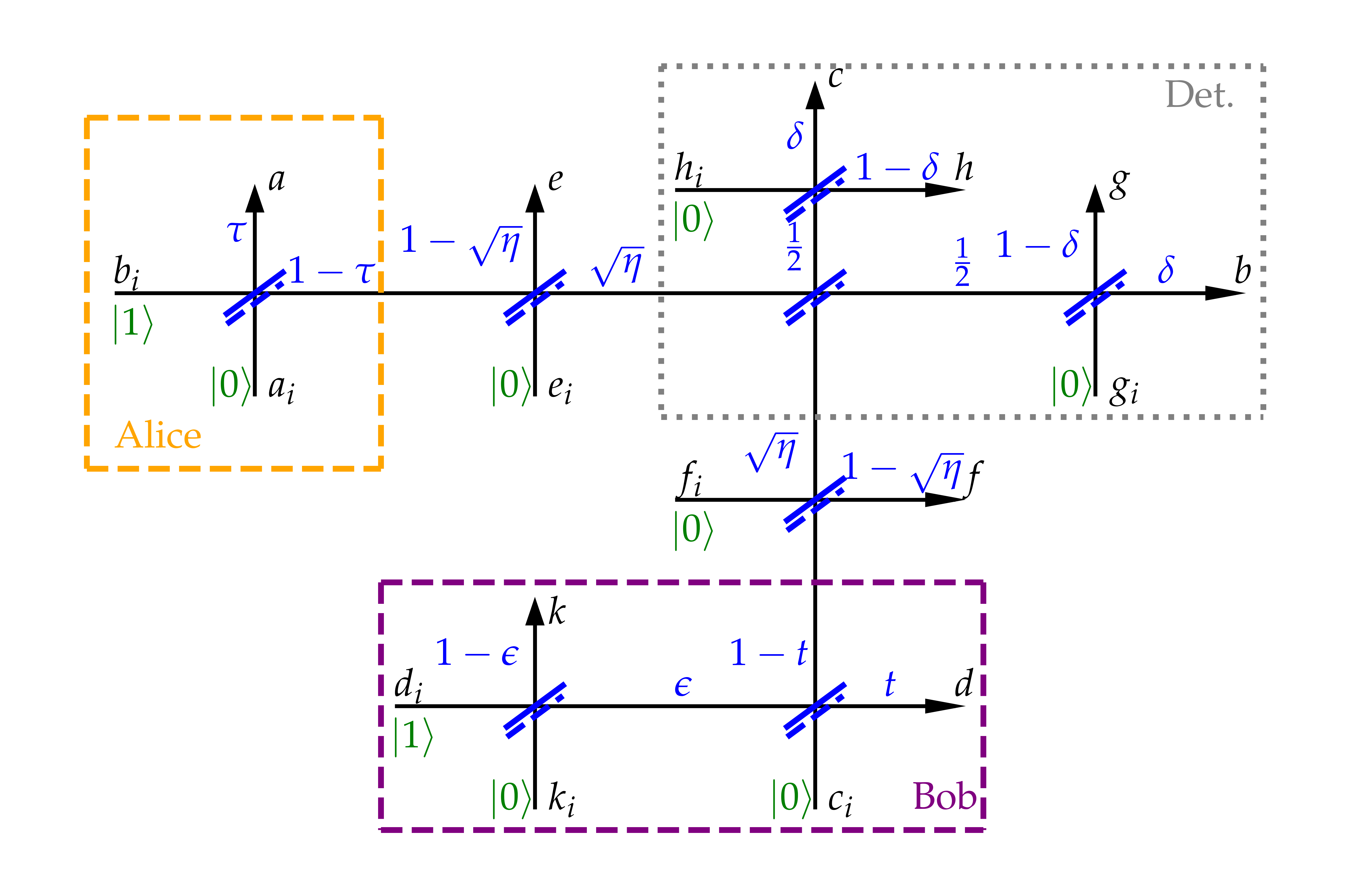}}
\caption{NLA protocol in presence of imperfect source and detection components. Detection noise is modelled as in the previous case. Source quality is modelled in a similar way, with a probability $1-\epsilon$ for the photon to get lost. We assume Alice has a perfect entangled state in order to focus the analysis on the noise introduced by the protocols.}
\end{figure}

\subsection{NLA at Bob's end}

For this case represented in Figure \ref{diag5}, the click probability is 
 \begin{equation}
P_{\text{success}}    = \epsilon \delta \tau  \left(1-t \right) + \epsilon \delta \left( 1- \tau \right) \left( 1 - \eta \right)  \left( 1 - t \right)  + \epsilon \delta \left( 1 - \tau \right) \eta  t  + 2 \epsilon ( 1- \tau) (1-t) (1- \delta) \eta \delta + \delta (1-\tau) \eta (1-\epsilon)
\label{psuccess_end_meas_source_noise}
            \end{equation}
  The projection of a click state $\vert \psi \rangle$ onto no environmental losses gives
       \begin{equation}
 P_f =  \frac{1}{2}  \tau \delta \epsilon \left(1-t \right) +  \frac{1}{2}  \left( 1-\tau \right) \delta \eta t \epsilon
     \end{equation}
       and projecting onto environmental losses gives
          \begin{equation}
P_0 =  \delta \epsilon (1-\tau) (1-t) \left( \frac{1}{2} (1-\eta) + \eta ( 1-\delta) \right) +  \frac{1}{2} \left(1-\tau \right)  \eta \delta (1-\epsilon)
\end{equation}
The state $  \vert \psi_{f} \rangle $ can be written in the same way as in section \ref{app:nla-perfect-ref}, with only a factor of $\sqrt{\delta} \sqrt{\epsilon} $ in front of it. The condition on $\tau$ to reach maximum entanglement is therefore unchanged
          \begin{equation}
  \tau = \frac{t \eta}{1 - t + t \eta}
  \end{equation}
With this condition, the click probability is given by
       \begin{equation}
    P_{\text{success}} = 2 \delta \epsilon \eta \frac{ t(1-t)}{1-t +t \eta} + \epsilon \delta (1+ \eta - 2 \eta \delta) \frac{ (1-t)^2}{1-t+t \eta} + \eta \delta (1-\epsilon) \frac{1-t}{1-t+t \eta}
    \label{interm_orig_meas_source_noise}
    \end{equation}
    The ratio $X$ representing state purity is given by
         \begin{equation}
       X  = \frac{ 2 t \eta  \epsilon}{  \epsilon (1+\eta - 2 \eta \delta)(1-t) + \eta(1-\epsilon)}
    \label{x_eq_orig_meas_noise}
          \end{equation} 
Here we note that an arbitrarily high $X$ cannot be reached anymore by setting $t$ closer and closer to 1, and the maximum $X$ is given by  
         \begin{equation}
      X_{\text{max}} =   \frac{2 \epsilon}{1-\epsilon},
      \end{equation}
      a level at which the click probability is exactly zero.
  Equation \ref{x_eq_orig_meas_noise} can still be inverted to obtain $t$ as a function of $X$, such that       
     \begin{equation}
t = \frac{X \left( \epsilon ( 1 + \eta - 2 \eta \delta) +\eta ( 1 - \epsilon) \right) }{2 \eta  \epsilon + X   \epsilon ( 1 + \eta - 2 \eta \delta)  }
\end{equation}
After replacing all occurrences of $t$ by their expressions in Equation \ref{interm_orig_meas_source_noise} and some simplifications, we get
   \begin{equation}
\boxed{
    P_{\text{success}} =    \frac{  2 \delta  \eta  (1+X) (\epsilon + \eta - 2 \epsilon \eta \delta )  \left( 2  \epsilon - X  ( 1 - \epsilon)  \right)}  {\left[ 2 \eta   + X   ( 1 + \eta - 2 \eta \delta) \right] \left[2 \epsilon +  X  \left( 2 \epsilon (1 - \eta \delta) + \eta -1 \right)    \right] } 
   }
   \label{psuccess_final_orig_meas_source_noise}
 \end{equation}
 If we assume $X \gg 1$, $\eta \ll 1$ and $1-\epsilon \ll 1$ (ie, very good source quality), the expression reduces to
   \begin{equation}
    P_{\text{success}} \approx 4 \delta \eta \left( \frac{1}{X} - \frac{1-\epsilon}{2} \right)
    \end{equation}

   \subsection{NLA half-way}

For this case represented in Figure \ref{diag6}, the click probability is given by
  \begin{equation}
P_{\text{success}}    = \delta \epsilon \sqrt{\eta} \left( (1- \tau) t + \tau ( 1-t) \right) + 2 \delta \epsilon \sqrt{\eta} (1-\tau)(1-t) \left(1- \delta \sqrt{\eta}  \right)+ (1-\tau)\sqrt{\eta} \delta (1-\epsilon)
    \label{interm_mid_meas_source_noise}
 \end{equation}
  The projection of a click state $\vert \psi \rangle$ onto no environmental losses gives
   \begin{equation}
 P_f = \frac{1}{2} \sqrt{\eta} \delta \epsilon \left( (1-\tau) t + \tau ( 1-t) \right)
 \end{equation}
  and projecting onto environmental losses gives
     \begin{equation}
P_0 =  \epsilon (1-\tau)(1-t)\delta \sqrt{\eta} (1-\delta \sqrt{\eta} ) + \frac{1}{2} (1-\tau) \delta (1-\epsilon)\sqrt{\eta}
\end{equation}
The state $  \vert \psi_{f} \rangle $ can be written in the same way as in section \ref{app:nla-perfect-ref}, with only a factor of $\sqrt{\delta} \sqrt{\epsilon} $ in front of it. The condition on $\tau$ to reach maximum entanglement is therefore unchanged
          \begin{equation}
  \tau = t
  \end{equation}
    With this condition, the click probability is given by
    \begin{equation}
    P_{\text{success}}    = 2 \delta \epsilon \sqrt{\eta} t (1- t) + 2 \delta \epsilon \sqrt{\eta} (1-t)^2 \left(1- \delta \sqrt{\eta}  \right)+ (1-t)\sqrt{\eta} \delta (1-\epsilon)
    \end{equation}
    The ratio $X$ representing state purity is given by
    \begin{equation}
       X  =  \frac{\epsilon t}{\epsilon (1-t) (1-\delta \sqrt{\eta} ) + \frac{1}{2} (1-\epsilon)}
    \label{x_eq_mid_meas_source_noise}
          \end{equation} 
          Similarly to the previous case, an arbitrarily high $X$ cannot be reached and the maximum $X$ is given by
         \begin{equation}
      X_{\text{max}} =   \frac{2 \epsilon}{1-\epsilon},
      \end{equation}
      a level at which the click probability is exactly zero.     The formula can still be inverted to obtain $t$ as a function of $X$, such that    
\begin{equation}
t = \frac{X \left[ \epsilon (1-\delta \sqrt{\eta} ) + \frac{1}{2} (1-\epsilon) \right] }{\epsilon  + X \epsilon (1-\delta \sqrt{\eta}) }
\end{equation}   
After replacing all occurences of $t$ by their expressions in Equation \ref{interm_mid_meas_source_noise} and some simplifications, we get
   \begin{equation}
\boxed{
P_{\text{success}}    = 
    \frac{\delta  \sqrt{\eta}  (1+X) \left( 1 + \epsilon   -   2 \epsilon \delta \sqrt{\eta}   \right)  \left( 2 \epsilon  -  X(1-\epsilon) \right) }{2 \epsilon  \left(1 + X   (1-\delta \sqrt{\eta}) \right)^2 }
    }
   \label{psuccess_final_mid_meas_noise}
 \end{equation}
 If we assume $X \gg 1$, $\eta \ll 1$ and $1-\epsilon \ll 1$ (ie, very good source quality), the expression reduces to
   \begin{equation}
    P_{\text{success}} \approx 2 \delta \sqrt{\eta} \left( \frac{1}{X} - \frac{1-\epsilon}{2} \right)
    \end{equation}
     
\section{Purification Protocol}
\label{sec:app-purif}

\subsection{Perfect Source and Detection}

From the diagram in Figure \ref{diagram_purif}, the input state for the entire system is 
\begin{equation}
   \vert 010100 \rangle_{a_i b_i c_i d_i e_i f_i} \otimes \vert \Omega_A \rangle \otimes \vert \Omega_B \rangle = b_i^\dagger d_i^\dagger \vert 000000 \rangle _{a_i b_i c_i d_i e_i f_i} \otimes \vert \Omega_A \rangle \otimes \vert \Omega_B \rangle
   \end{equation}
   with
        \begin{align}
  \begin{split}
   \vert \Omega_A \rangle &= \sqrt{\frac{1}{2}} g_i^\dagger  \vert 000 \rangle _{k_i h_i g_i} + \sqrt{\frac{1}{2}} k_i^\dagger h_i^\dagger   \vert 000 \rangle _{k_i h_i g_i} \\
 \vert \Omega_B \rangle &=  \sqrt{\frac{1}{2}} m_i^\dagger \vert 000 \rangle _{m_i n_i p_i} + \sqrt{\frac{1}{2}} n_i^\dagger p_i^\dagger   \vert 000 \rangle _{m_i n_i p_i} 
    \end{split}
   \end{align}
   
The relevant operators can be written as a function of all the output operators
       \begin{align}
  \begin{split}
   b_i^\dagger &= \sqrt{1-\tau} \sqrt{\frac{1}{2}} b^\dagger +  \sqrt{1-\tau}\sqrt{\frac{1}{2}} k^\dagger 
   - \sqrt{\tau} \sqrt{\eta} \sqrt{\frac{1}{2}} a^\dagger -  \sqrt{\tau} \sqrt{\eta} \sqrt{\frac{1}{2}} p^\dagger
   - \sqrt{\tau} \sqrt{1-\eta} f^\dagger \\
   d_i^\dagger &= \sqrt{1-t} \sqrt{\frac{1}{2}} d^\dagger -  \sqrt{1-t}\sqrt{\frac{1}{2}} g^\dagger 
   + \sqrt{t} \sqrt{\eta} \sqrt{\frac{1}{2}} c^\dagger -  \sqrt{t} \sqrt{\eta} \sqrt{\frac{1}{2}} m^\dagger 
   -    \sqrt{t}  \sqrt{1-\eta} e^\dagger \\
   k_i^\dagger &= - \sqrt{\frac{1}{2}} b^\dagger +  \sqrt{\frac{1}{2}} k^\dagger \\
   g_i^\dagger &= \sqrt{\frac{1}{2}} d^\dagger +  \sqrt{\frac{1}{2}} g^\dagger \\ 
   m_i^\dagger &= \sqrt{\frac{1}{2}} c^\dagger +  \sqrt{\frac{1}{2}} m^\dagger \\
   p_i^\dagger &= - \sqrt{\frac{1}{2}} a^\dagger +  \sqrt{\frac{1}{2}} p^\dagger \\
   h_i^\dagger &= h^\dagger \\
   n_i^\dagger &= n^\dagger
   \end{split}
   \end{align}
After expanding the input state and projecting onto no environment losses (with the help of some computer code), we obtain \begin{equation}
   \begin{split}
 \vert  \psi_f \rangle &= - \frac{1}{8} \sqrt{\eta} \sqrt{1-t} \sqrt{\tau} \vert 10 \rangle_{hn} +  \frac{1}{8} \sqrt{\eta} \sqrt{t} \sqrt{1-\tau} \vert 01 \rangle_{hn}  \\
   P_f &= \frac{1}{64} \eta ( \tau(1-t) + t (1-\tau) )
   \end{split}
   \end{equation}
Projecting onto environmental loss states doesn't yield any term such that $P_0=0$. The condition of maximum entanglement yields $\tau=t$, from which we obtained Equation \ref{purif_psuc_perfect}.

\subsection{Imperfect Source and Detection}

Following the same method, we get a click probability of
    \begin{equation}
   P_{\text{success}} = \frac{1}{4} \delta^4 \epsilon^2 \eta ( t (1-\tau) + \tau(1-t) )
   \end{equation}
 A click state without environmental loss is
   \begin{equation}
   \begin{split}
 \vert  \psi_f \rangle &= - \frac{1}{8} \delta^2 \epsilon \sqrt{1-t} \sqrt{\tau} \sqrt{\eta} \vert 10 \rangle_{hn} +  \frac{1}{8} \delta^2  \epsilon^2 \sqrt{\eta} \sqrt{t} \sqrt{1-\tau} \vert 01 \rangle_{hn}  \\
   P_f &= \frac{1}{64} \delta^4 \epsilon^2 \eta ( \tau(1-t) + \epsilon^2 t (1-\tau) )
   \end{split}
   \end{equation}
 and a state with environmental loss $\vert \psi_0 \rangle$ has probability
    \begin{equation}
   P_0 = \frac{1}{64} \delta^4 \epsilon^2 \eta t (1 - \epsilon) (1 - \tau) (1 + \epsilon)
   \end{equation}
 (and we verify $P_{\text{success}} = 16 (P_0 + P_f)$ for 16 successful states). 
  The condition of max entanglement for $\vert \psi_f \rangle$ gives
 \begin{equation}
 \tau = \frac{\epsilon^2 t}{1 - t + \epsilon^2 t}
 \end{equation}
 We can calculate $X$ and $P_{\text{success}}$ now with this condition and we get
 \begin{equation}
 X = \frac{2 \epsilon^2}{1-\epsilon^2} \quad\mathrm{and}\quad  P_{\text{success}} = \frac{ \delta^4 \epsilon^2 \eta (1+\epsilon^2)}{4 (1+\epsilon)^2} 
 \end{equation}
 If we assume $1-\epsilon \ll 1$, the expression for  $P_{\text{success}}$ reduces to 
  \begin{equation}
   P_{\text{success}} \approx \frac{ \delta^4 \epsilon^2 \eta}{8}
   \end{equation}
 
%\bibliography{bib1}

%apsrev4-2.bst 2019-01-14 (MD) hand-edited version of apsrev4-1.bst
%Control: key (0)
%Control: author (8) initials jnrlst
%Control: editor formatted (1) identically to author
%Control: production of article title (0) allowed
%Control: page (0) single
%Control: year (1) truncated
%Control: production of eprint (0) enabled
%

\end{document}